%% file: main.tex
\documentclass[a4paper,11pt]{article}
\usepackage{jheppub} 
\usepackage{physics}
\usepackage{subfigure}
\usepackage{subfigure}
\usepackage{ascmac}
\usepackage{amsthm}
\usepackage{amsmath}
\usepackage{tikz}
\usepackage{tikzit}
\input{fig/string_scattering.tikzstyles}
\usetikzlibrary{arrows.meta}
\usetikzlibrary{patterns}
\usetikzlibrary{decorations.pathmorphing}
\usetikzlibrary{decorations.text}
\usetikzlibrary{snakes}
\tikzset{>={Latex[width=2mm,length=2mm]}}

\allowdisplaybreaks

\newcommand{\Freg}[4]{
	\mathbf{F}\left(
	\begin{array}{cc}
		#1, & #2 \\
		\multicolumn{2}{c}{#3}
	\end{array}; #4
	\right)
}

\preprint{KUNS-2939}

\title{
    Transient chaos analysis of string scattering
}

\author{Koji Hashimoto$^{*}$, Yoshinori Matsuo$^{\dagger}$, Takuya Yoda$^{*}$}
\affiliation{${}^*$Department of Physics, Kyoto University, Sakyo-ku, Kyoto 606-8502, Japan}
\affiliation{${}^\dagger$Department of Physics, Kindai University, Higashi-Osaka, Osaka 577-8502, Japan}

\emailAdd{koji@scphys.kyoto-u.ac.jp}
\emailAdd{ymatsuo@phys.kindai.ac.jp}
\emailAdd{t.yoda@gauge.scphys.kyoto-u.ac.jp}

\abstract{
    It has long been thought that
    a highly excited string can be regarded as a black hole:
    the correspondence principle between strings and a black hole,
    while recent studies found that
    black holes are characterized by chaos.
    This suggests that
    highly excited strings are the source of the black hole chaoticity.
    We study
    the chaoticity of a string amplitude
    where a tachyon is scattered by a highly excited string.
    Our strategy to extract the chaos in the amplitude
    is a generalization of the transient chaos analysis for classical scattering.
    We look for the fractal structure
    in the plots of incoming/outgoing scattering angles,
    where the outgoing angle is defined as the maximum pole of the amplitude.
    Within our strategy,
    we could not identify any fractal structure in the scattering data.
    We also discuss
    other possible setups and strategies
    to extract the chaos,
    hoping that
    our present work serves as a step toward the formulation of chaos in string scattering amplitudes.
}

\begin{document} 
\maketitle
\flushbottom

\section{Introduction and summary}

It has long been thought that a highly excited string can be regarded as a black hole: the correspondence principle between strings and a black hole \cite{Horowitz:1996nw,Horowitz:1997jc,Amati:1999fv,Chen:2021dsw}.
The correspondence stems naturally from the fact that string theory is a quantum gravity, and that a black hole is a gravitational phenomenon in which many particles, that are strings in string theory, gather to form a ``bound state." In addition, this correspondence is motivated by the effort to reveal the mystery of the black hole entropy formula, since any explicit formulation of the correspondence may lead to the way to count the quantum states of a black hole.

In recent years, black holes were characterized by chaos, in the AdS/CFT correspondence \cite{Maldacena:1997re}. The chaotic nature of the boundary quantum field theory corresponds to the redshift due to the black hole event horizon in the gravity side \cite{Shenker:2013pqa,Shenker:2013yza,Maldacena:2015waa}. In view of this development, chaos can be one of the best ways to confirm the correspondence principle between the highly excited string and the black hole.

Since perturbative string theory is formulated through scattering amplitudes, we need to look for any chaotic nature of the string scattering amplitudes. Recently, Gross and Rosenhaus \cite{Gross:2021gsj} studied the amplitudes of a highly excited string decaying into two tachyons, and found that the amplitudes are highly erratic, which is a sign of chaos. The erratic behavior was first mentioned in \cite{Rosenhaus:2020tmv} and further confirmed in \cite{Rosenhaus:2021xhm}.\footnote{See also \cite{Fukushima:2022lsd,Firrotta:2022cku,Bianchi:2022mhs} for related recent work on scattering and chaos in string theory.}
In addition, in our recent work \cite{Hashimoto:2022ugt}, the imaging of the highly excited string through the decay amplitudes revealed the nontrivial spatial structure of the string: string is a double slit. The existence of such a structure is a sign of chaos.

In order to extract the chaos in the erratic behavior of the decay amplitude of the highly excited string,
we resort to an established method: the transient chaos analysis. In classical scattering, the diagnose
for finding chaos is to identify a fractal structure in the scattering data (see \cite{seoane2012new} for a review).
The methods of the transient chaos analysis has been developed such that the chaos of bounded systems is generalized
to unbounded systems --- in the scattering processes, the chaos appears only in a limited spatial and temporal domain,
while for the bounded systems one can observe the chaos using infinite amount of time.

The transient chaos analysis which we employ in this paper is summarized as follows.
Consider a particle scattered by a potential. The scattering data consists of a pair of the incoming and outgoing angles
of the particle motion, and we shall name it $(\theta, \theta')$. For a sufficient number of numerical experiments of the scattering, one finds a lot of the pair data, which is translated into a one-dimensional function $\theta'(\theta)$. For chaotic
scattering, one should be able to find a fractal structure in this function $\theta'(\theta)$.
Like in the chaotic dynamics produced by the Baker's map, the fractal structure or the self-similar structure is a typical nature of chaos. In scattering processes, if there exists the initial condition sensitivity in the system, a particle destination is shared by two or more different paths. This produces the fractal structure in the scattering data.

In this paper, we apply the transient chaos analysis described above to string scattering amplitudes. In particular, we consider scattering amplitudes of a tachyon and a highly excited string (HES), {\it i.e.}\ amplitudes of HES-tachyon to HES-tachyon, which mimic a scattering of a tachyon by a HES. To describe the in- and out-HES state, we follow the method developed in \cite{Gross:2021gsj} using the DDF states \cite{DelGiudice:1971yjh}. 

To obtain the scattering data $\theta'(\theta)$ from the quantum scattering amplitudes, we extract
the outgoing angles of the largest pole $\theta'$ of the amplitude for fixed incoming angles $\theta$. Since generically string scattering amplitudes have multiple poles, we regard the largest pole as the most probable scattering in the quantum treatment.

Our final results of the scattering data of HES-tachyon scattering in open bosonic string theory are shown in Fig.~\ref{fig:theta_thetap_J=1}, 
Fig.~\ref{fig:theta_thetap_J=1_higher} and Fig.~\ref{fig:theta_thetap_J=gen}. The HES states are the ones up to the string excitation level $N=27$, and the photon insertions necessary for creating the HES state are taken up to $J=5$, and we consider generic out-going angles. 
We do not see any fractal structure in the scattering data, which concludes that our HES-tachyon scattering analysis does not show the sign of transient chaos.

One caveat is that the absence of the fractal structure may originate in the quantum nature of the string scattering amplitudes. As described above, our strategy to obtain the scattering data is 
to pick the largest pole in the amplitude. There could be a refined strategy to extract the scattering data by taking some classical limit of the scattering amplitudes.

This paper is organized as follows. In Sec.~\ref{sec:review}, we make a brief review of the transient chaos, and propose how the scattering data is extracted from the quantum scattering amplitudes,  which is used in our analysis.
In Sec.~\ref{sec:HES-t}, we calculate the HES-tachyon scattering amplitudes for various HES states and obtain the scattering data, to show that there is no fractal structure in the scattering data of the amplitudes.
In Sec.~\ref{sec:conclusion}, we provide discussions on possible generalization of our method to find chaos in string scattering amplitudes. 
App.~\ref{sec:hes-tachyon_amp} is for the detailed derivation of the HES-tachyon scattering amplitudes used in Sec.~\ref{sec:HES-t}.

\section{Transient chaos in classical scattering}
\label{sec:review}

In this section,
we briefly review the fractal feature of transient chaos
in classical scattering processes, and propose
its generalization to quantum scattering processes, for our study of string scattering in the later sections.

\subsection{Classical scattering and chaos}

The chaos in classical scattering processes can be identified by finding a fractal structure in the scattering data. 
We briefly review how the fractal nature appears in the classical scattering, following \cite{seoane2012new}.

\begin{figure}[t]
	\centering
\includegraphics[keepaspectratio, scale=0.5]{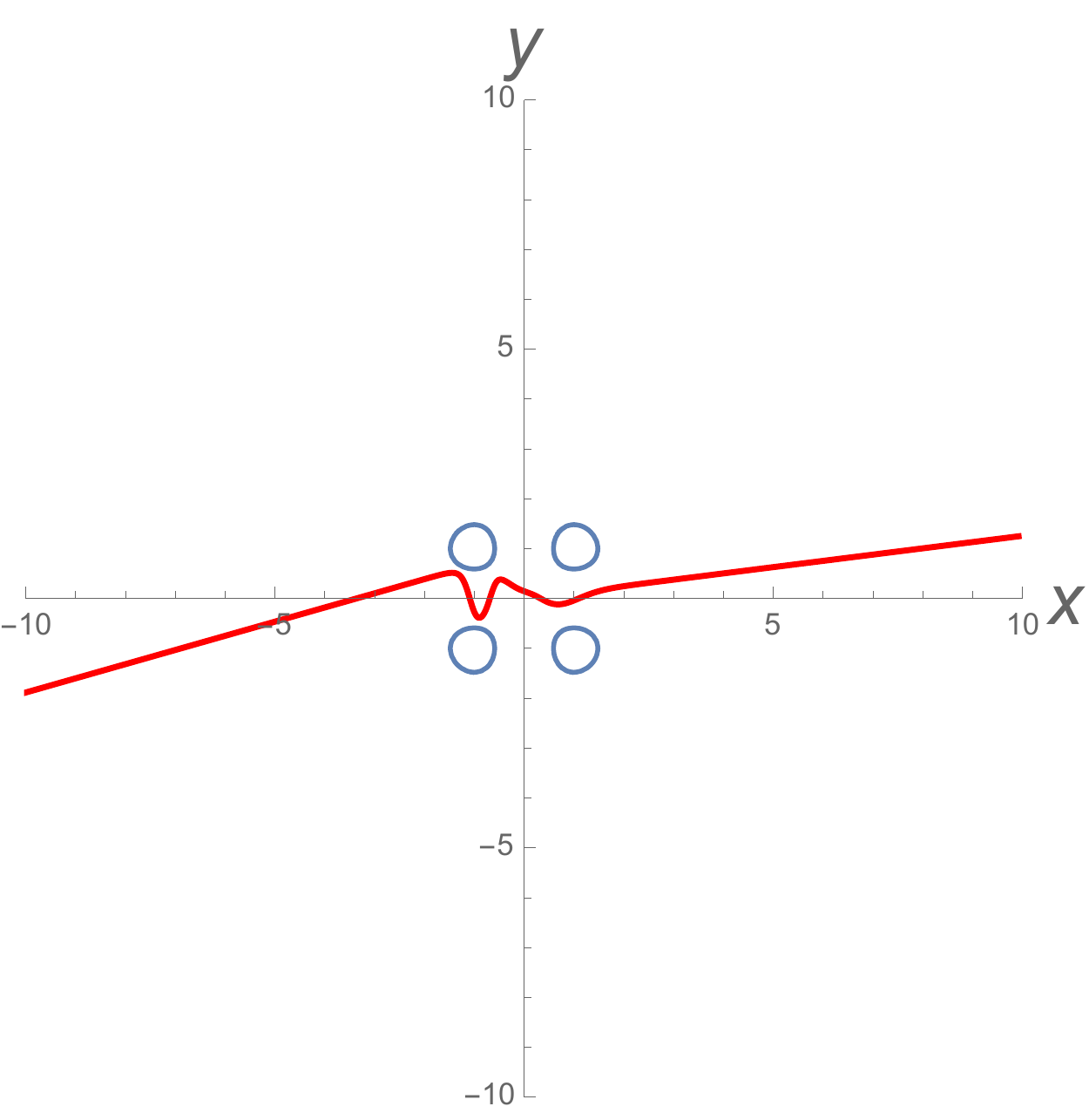}
\includegraphics[keepaspectratio, scale=0.5]{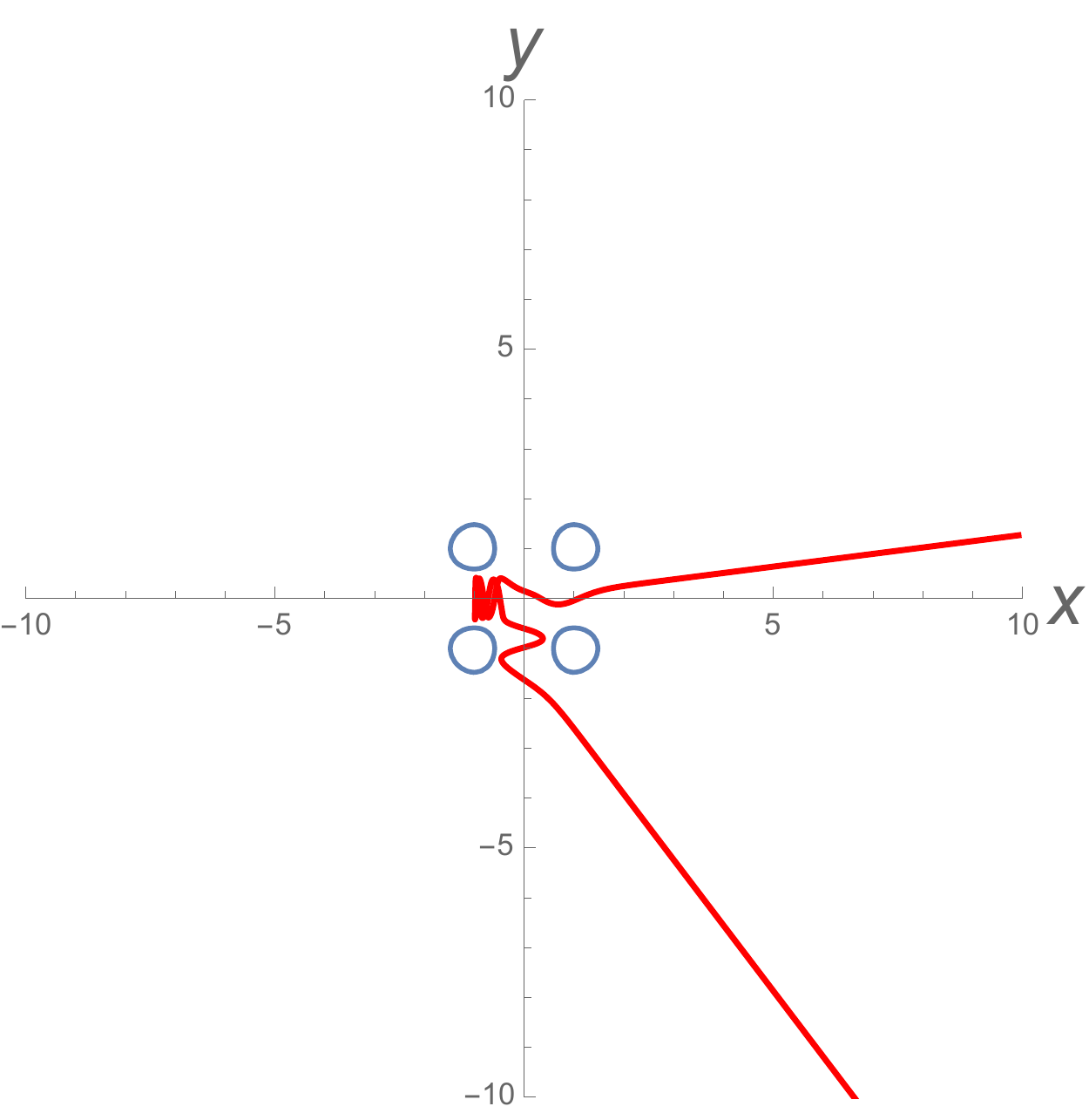}
\caption{The classical scattering of a single particle in two spatial dimensions. The scattering potential is the four-hill potential \eqref{four-hill} whose equal-height slice is plotted as four circles. Left: Initial condition is $\theta=0.125$ with $b=0$. Right: Initial condition is $\theta=0.127$ with $b=0$. For both plots, the energy is fixed to $E=0.045$. }
	\label{fig:scatter}
\end{figure}

Consider the motion of a single particle scattered by a potential (see Fig.~\ref{fig:scatter}). A potential is localized around the origin of the space into which the particle is shot and then scattered to reach the spatial infinity.
The scattering data could be, for example, the incoming angle $\theta$ and the impact parameter $b$ for the initial state of the particle and the outgoing angle $\theta'$ and the impact parameter $b'$ for the final state of the particle. Thus the data consists of a set of values $(\theta, b, \theta', b')$ with many numerical experiments, for a fixed potential and fixed energy of the particle. In other words, the data is a set of functions $\theta'(\theta,b)$ and $b'(\theta,b)$.\footnote{One can choose $\theta=0$ and ignore measuring $b'$, then the scattering data is only $\theta'(b)$, which was used in \cite{bleher1990bifurcation}.}

If one chooses a potential whose complexity is large enough, then it happens that the particle which was shot in
can be trapped for a certain period of time and then escapes to the spatial infinity. During this trapping time,
the particle is scattered many times by the inner structure of the potential and loses its initial information (which is $\theta$ and $b$). This causes chaos, the sensitivity to the initial condition.

In Fig.~\ref{fig:scatter}, 
two examples of the scattering of a particle which goes through the four-hill potential with the Lagrangian
\begin{align}
    L = \frac12 \dot{x}^2 + \frac12 \dot{y}^2-x^2 y^2 e^{-x^2-y^2}
    \label{four-hill}
\end{align}
are presented. The particle is shot from the right hand side with the initial angle $\theta=0.125$ and $0.127$
respectively, with $b=0$. The outcome is drastically different, although the initial conditions differ just slightly.

Generic scattering processes are different from the case of the bounded systems for which the standard definition of
chaos applies: one cannot measure the chaos for infinite duration of time, because the particle can escape from the chaotic region. Therefore the chaotic nature is hidden in the scattering data of the particle which goes through the potential.

\begin{figure}[t]
	\centering
\includegraphics[keepaspectratio, scale=0.3]{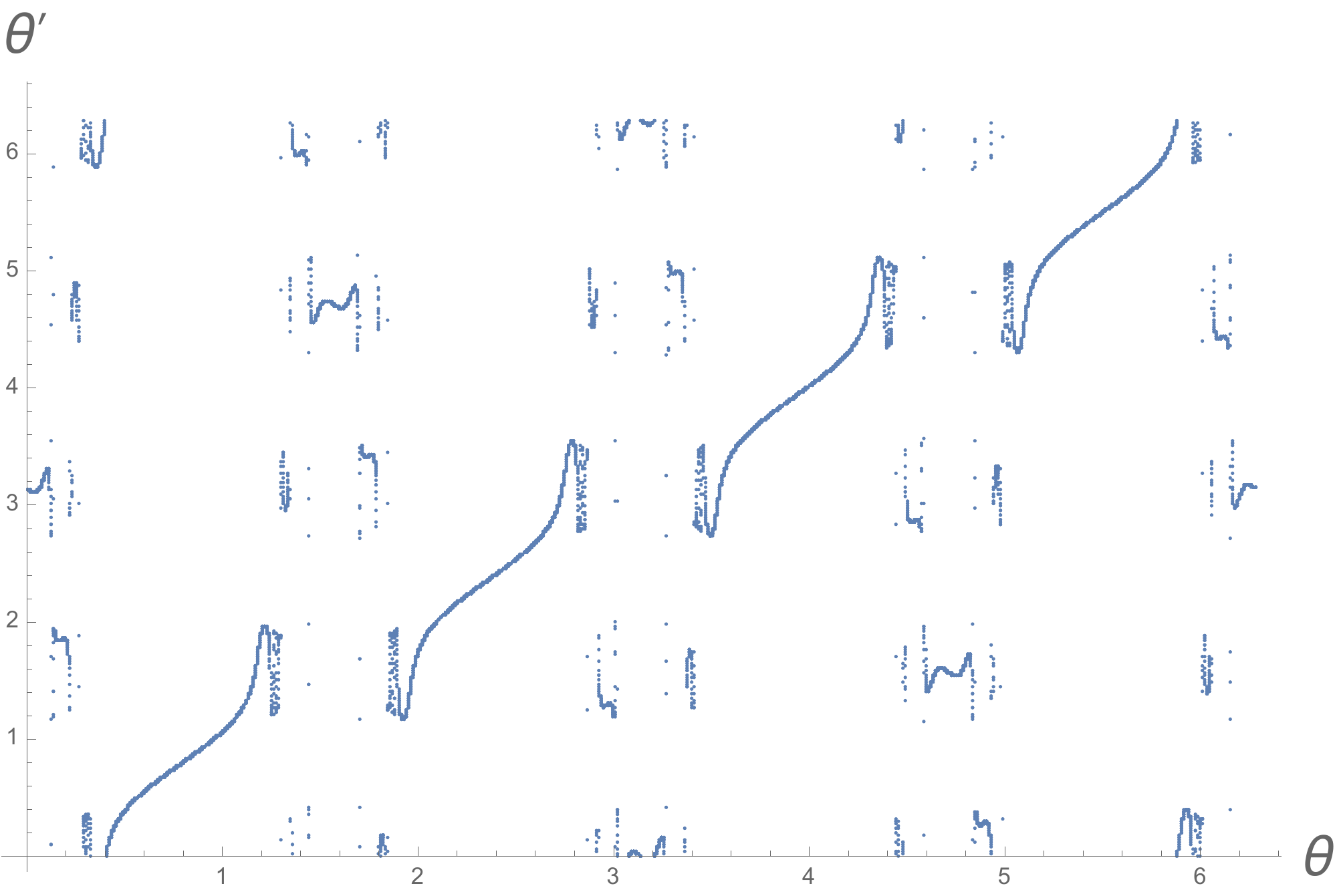}
\hspace{5mm}
\includegraphics[keepaspectratio, scale=0.3]{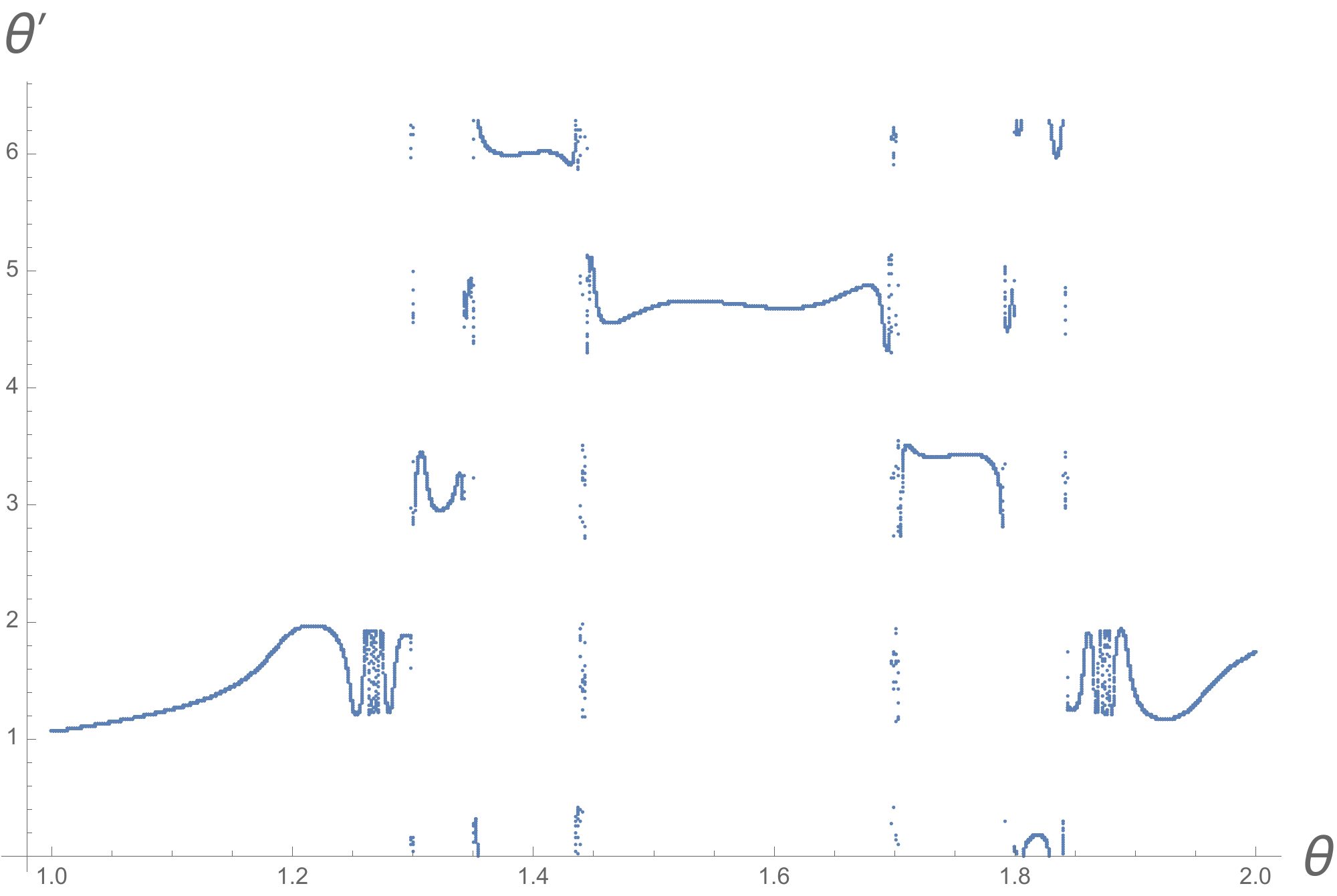}
\\
\includegraphics[keepaspectratio, scale=0.3]{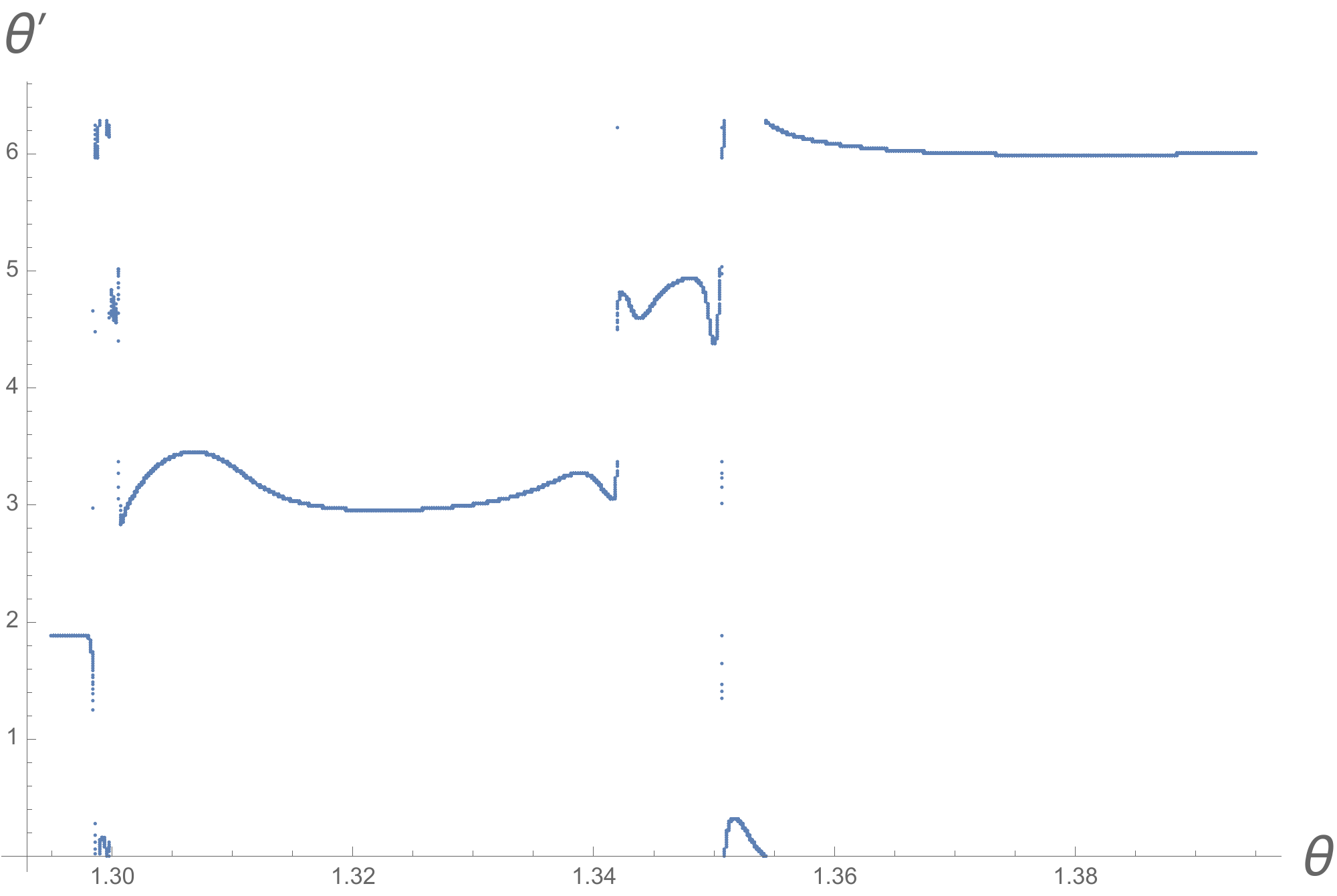}
\hspace{5mm}
\includegraphics[keepaspectratio, scale=0.3]{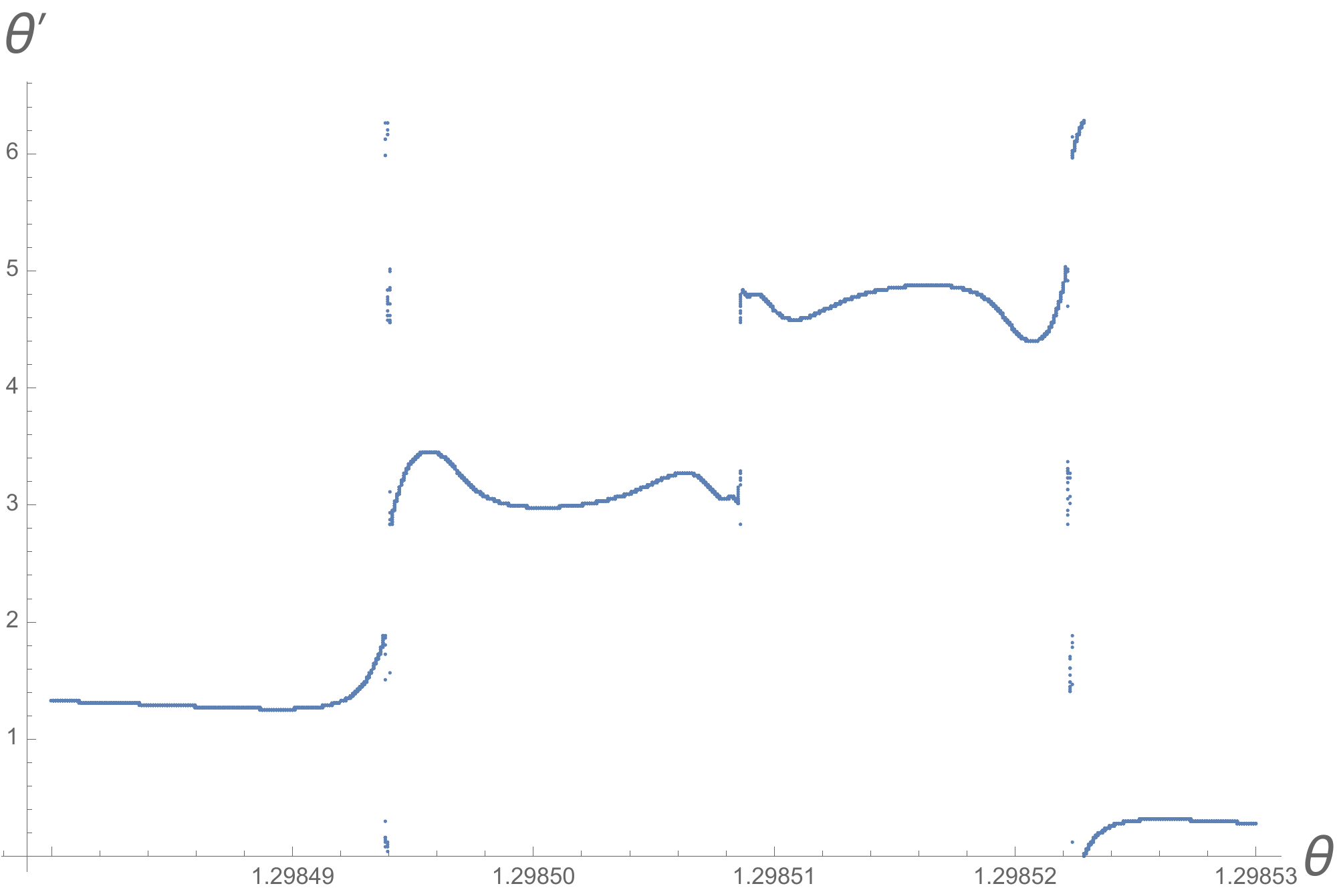}
\caption{The scattering data of the particle for the four-hill potential model \eqref{four-hill}. The particle energy is fixed to $E=0.045$, and the impact parameter is also fixed to $b=0$. Top-left: the whole view of the scattering data $\theta'(\theta)$. The top-right, bottom-left, bottom-right panels are the magnified plots of the scattering data. One can find the fractal-like structure of the self-similarity.}
	\label{fig:scattering_data}
\end{figure}

In Fig.~\ref{fig:scattering_data}, the scattering data of the particle which goes through the four-hill potential
\eqref{four-hill} is presented. We have chosen $b=0$ for simplicity, and look at only the function $\theta'(\theta)$.
The top-left panel is the whole view of the data. The scattering data obviously finds a lot of jumps, and in particular, the jumps are accumulated to form a dense part. If one magnifies the dense part, as seen in the other panels, one finds a similar dense structure in the scattering data. This is the fractal-like structure, which shows the transient chaos of the scattering process.

The existence of the fractal structure means that a single destination of the particle motion can have two or more paths. It is nothing but the initial condition sensitivity, that is the definition of classical chaos.

In summary, for classical scattering, the fractal structure in the scattering data $\theta'(\theta)$ shows the chaos. In this paper, we obtain the scattering data for quantum scattering of strings, in particular highly excited strings, and look for the chaos, that is, the fractal structure.

\subsection{Generalization to quantum scattering}
\label{sec:gen_quantum_scattering}

The transient chaos in classical scattering is well-understood, as we have reviewed in the previous subsection. However, string theory is built out of quantum scattering amplitudes. How can we generalize the classical
analysis to the quantum scattering amplitudes?
Here, let us explain our strategy to apply the
transient chaos analysis for classical scattering processes which we reviewed in the previous subsection to the quantum scattering processes.

First of all, we need to note that in quantum scattering processes, due to the Heisenberg uncertainty principle, we do not have all the information which the classical scattering has.
Thus here we start with a discussion on 
what parameters are suitable for the quantum scattering processes.

The quantum scattering in standard quantum field theories is normally described by plane waves, as any particle state is a superposition of quantized plane waves.
This means that, the initial state and the final state are specified only by the momenta of all the particle states, incoming and outgoing.
They are the parameters for quantum scattering, and we can further reduce the number of the independent parameters by considering some limit, as follows.

In the analogy of the physical set-up of the transient chaos in classical scattering, we suppose a scattering of a light particle by a potential. 
In the standard 2-to-2 scattering processes in quantum field theories, this potential scattering is realized if one requires a limit where the other particle is much heavier than the light scattered particle. Thus, in this limit, what specifies the scattering amplitude is only the combination 
$(p_\mu^{\rm (ini)},p_\mu^{\rm (fin)})$,
which is the momenta of the light scattered particle.
Furthermore, once we specify the mass $m$ of the light particle, then with the energy conservation, we find
$p_0^{\rm (ini)}=p_0^{\rm (fin)}=\sqrt{p^2 + m^2}$ for the momentum magnitude $p$ ($>0$). The scattering is specified by $p$ and the pair $(\theta, \theta')$
where $\theta$ and $\theta'$ are the angles of the incoming and the outgoing momenta $(p_i^{\rm (ini)}, p_i^{\rm (fin)})$. Thus, the scattering data is given by the scattering amplitude ${\cal A}(p,\theta, \theta')$.

Next, we present our procedures to extract the scattering plot similar to Fig.~\ref{fig:scattering_data} from the 
quantum scattering amplitudes, to find the transient chaos of the system. 
Note that the classical scattering data is given by a map $(b, \theta) \to (b', \theta')$ for a given energy, where $b$ is the impact parameter. In the quantum scattering we consider plane waves, meaning that the impact parameters are uniformly integrated.
Once the impact parameter dependence is integrated, the classical scattering is specified by a map $\theta \to \theta'$, {\it i.e.} a function $\theta'(\theta)$
for a given energy. 

We can relate this classical scattering data $\theta'(\theta)$ and the quantum scattering amplitude ${\cal A}(p,\theta, \theta')$ as follows.
The absolute square of the scattering amplitude ${\cal A}$ describes the probability.
When the amplitude has a pole at some value of $\theta'$, we can say that the value is
most probable in the scattering. 
Since the position of the pole depends also on the other data $(p,\theta)$, 
it gives $\theta'$ as an implicit function of $(p,\theta)$. 
When the amplitude has multiple poles at $\theta'=\theta'_m$ for given $p$ and $\theta$, as
\begin{align}
    {\cal A}(p,\theta, \theta') \sim \sum_m \frac{R_m(p,\theta)}{\theta'-\theta'_m(p,\theta)}
\end{align}
then we can say that the value $\theta'_m(p,\theta)$ which gives the largest residue $|R_m(p,\theta)|$ among all poles is the most probable scattering.
In this manner, we can extract the function $\theta'(p,\theta)$, 
which is $\theta'_m(p,\theta)$ with the largest $|R_m(p,\theta)|$, 
from the scattering amplitude ${\cal A}(p,\theta, \theta')$, and regard 
$\theta'(p,\theta)$ as a quantum counterpart of the classical scattering data $\theta'(\theta)$. 

Note that the obtained function $\theta'(p,\theta)$ can be discontinuous, since the
largest pole, or equivalently the label $m$ of the largest residue, may differ for different $\theta$. In general, when there are multiple poles in the scattering amplitude, the scattering data $\theta'(\theta)$ for a given value of $p$, which is a single-valued, 
consists of a set of segment functions whose domains are disjoint with each other. 

In the following section, we use this definition of $\theta'(\theta)$ for the scattering data of the string scattering amplitudes.

\section{HES-tachyon to HES-tachyon scattering}
\label{sec:HES-t}

In this section,
we study a scattering of a tachyon
by a highly excited string (HES) in open bosonic string theory:
\begin{align}
    \label{eq:scattering_process_HES_tachyon}
    \text{HES} + \text{tachyon}
    \rightarrow
    \text{HES} + \text{tachyon}.
\end{align}
We compute its scattering amplitude $\mathcal{A}(p,\theta,\theta')$,
and find that
it has multiple poles at $\theta'=\theta_m'(\theta)$.
We regard
the angle associated with the largest pole
as a quantum counterpart of the classical scattering data.
We show that
the scattering data $\theta'(\theta)$
consists of disconnected segments,
and look for
the fractal structure.

\subsection{Setups}

\begin{figure}[t]
	\centering
	\input{fig/2_tachyon_J_photon.tikz}
	\caption{
		String amplitude for the process:
		(tachyon + $J$ photons) + tachyon $\rightarrow$ (tachyon + $J$ photons) + tachyon.
		The external lines on the left/right hand side
		are the incoming/outgoing strings.
		After picking out poles labeled with $v,v'$,
		it describes the scattering of a tachyon by a highly excited string (HES).
	}
	\label{fig:2_tachyon_J_photon}
\end{figure}
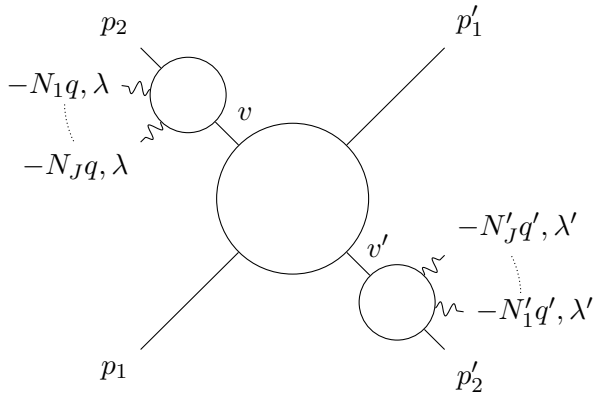

To construct the HES-tachyon scattering amplitude,
we start with a scattering:
\begin{align}
    (\text{tachyon} + J\: \text{photons}) + \text{tachyon}
	\rightarrow
	(\text{tachyon} + J\: \text{photons}) + \text{tachyon}.
\end{align}
Its diagram is shown in Fig.~\ref{fig:2_tachyon_J_photon}.
The external lines
on the left/right hand side
are the incoming/outgoing strings.
$J$ photons have momenta $\{ -N_aq \}_{a=1,\dots,J}$ and the same polarization $\lambda$ is shared by all the photons. 
The tachyon with momentum $p_2$ can form 
an intermediate HES state by absorbing all these $J$ photons.
By picking out a pole labeled with $v$,
we focus on the HES state.
A similar construction of a HES state
has been used, for example, in \cite{Rosenhaus:2021xhm,Gross:2021gsj}.
The HES scatters
another incoming tachyon with momentum $p_1$.
The outgoing strings are constructed in a similar but opposite processes.
The incoming strings decay into an outgoing tachyon with momentum $p_1'$
and a HES corresponding to a pole labeled with $v'$.
The outgoing HES state is constructed from a tachyon with momentum $p_2'$ 
and $J$ photons which have momenta $\{ -N_b'q' \}_{a=1,\dots,J}$ 
and share the same polarization $\lambda'$, 
or equivalently, the HES decays into a tachyon and $J$ photons successively.
The resulting amplitude
with the poles $v,v'$ extracted
describes the desired process \eqref{eq:scattering_process_HES_tachyon}.

Let us summarize the setups
and some assumptions to simplify computations.
The momentum conservation law is
\begin{align}
	0 &= p_1 + (p_2-Nq) + (p_2'-N'q') + p_1' \notag\\
	&= p_1 + p_2 + p_2' + p_1' - N(q+q') \ , 
\end{align}
where
\begin{align}
	N = \sum_{a=1}^{J} N_a = \sum_{b=1}^{J} N_b' = N', \quad
	(N_a, N_b' \geq 1)
\end{align}
is the total excitation level of HES, and here, 
the total excitation number $N'$ of the outgoing HES is taken to be 
the same as that ($N$) of the incoming HES. 
The mass $M$ of HES is given in terms of the total excitation number $N$ as 
\begin{align}
	M^2 = 2(N-1).
\end{align}
Since all photons have the same polarization, 
the number of photons $J$ equals to the total angular momentum.
Here note that
$N_b'$ is not necessarily equal to $N_b$.
The on-shell conditions for the tachyons and photons are
\begin{align}
	p_1^2 = p_2^2 = p_2'^2 = p_1'^2 = -(-2),
\end{align}
and 
\begin{align}
	q^2 = q'^2 = 0, 
\end{align}
respectively. 
For simplicity, we assume that $q\propto q'$ and $\lambda=-\lambda'$.
Then the photon momenta and polarization vectors satisfy
\begin{align}
	q\cdot q' &= 0,
\qquad\qquad
	\lambda^2 = \lambda'^2 = \lambda\cdot\lambda'= 0, 
\notag\\
	q\cdot \lambda &= q'\cdot\lambda' = q\cdot\lambda' = q'\cdot\lambda = 0.
\end{align}
We also assume that
\begin{align}
	(p_1+p_2)\cdot\lambda = (p_1'+p_2')\cdot \lambda' = 0
\end{align}
for simplicity.
This is satisfied without loss of generality,
for example,
by taking the center-of-mass frame of the incoming/outgoing strings.
We define the Mandelstam(-like) variables as
\begin{align}
	&s = -(p_1+p_2-Nq)^2 = -(p_1'+p_2'-N'q')^2, \\
	&t = -(p_1 + p_2'-N'q')^2 = -(p_1'+p_2-Nq)^2, \\
	&u = -(p_1+p_1')^2=-(p_2-Nq+p_2'-N'q')^2.
\end{align}
Also we define variables to describe the HES poles as
\begin{align}
	&v = -(p_2-Nq)^2, \\
	&v' = -(p_2'-N'q')^2.
\end{align}
These satisfy an identity
\begin{align}
	s+t+u
	= 4\cdot(-2) + 2\cdot(1+v/2)+2\cdot (1+v'/2).
	\label{Mandel-id}
\end{align}

\subsection{Amplitude and its largest pole}
Here, we calculate the HES-tachyon to HES-tachyon amplitude. 
The amplitude becomes a four point amplitude 
after picking out the intermediate HES poles of $v$- and $v'$-channels, 
and is given in the form of a worldsheet boundary integral over 
the position of one of four vertex operators 
which cannot be fixed by the conformal symmetry. 
The integral can be split into three segments. 
The amplitude contains poles of three channels: $s$-, $t$- and $u$-channels, 
and each of three segments of the integral contains 
contributions from two of three channels. 
Here, we focus on the segment of integral $\mathcal A_{st}$ 
which contains contributions of $s$- and $t$-channels.\footnote{The part $\mathcal A_{st}$ originates in the DDF construction of the fourth vertex operator going around one of the other three vertex operators, see App.~\ref{sec:hes-tachyon_amp} for the details. Although this is a part of the whole scattering amplitude, we assume that this is enough for looking at any possible fractal structure. In fact, the complete structure of the Veneziano amplitude is determined solely by $\mathcal A_{st}$.}

Some straightforward computations in App.~\ref{sec:hes-tachyon_amp} show that
the $st$-part contribution for
the HES-tachyon to HES-tachyon scattering amplitude is
\begin{align}
    \label{eq:amp_hes_tachyon_s-shifted_main}
	\mathcal{A}_{st}
	&\sim
	\sum_{ \{i_a=2,2'\} }
	\sum_{ \{j_b=2,2'\} }
	\sum_{ \{k_a=0\} }^{ \{N_a\} }
	\sum_{ \{l_b=0\} }^{ \{N_b'\} } \notag\\
	&\hspace{30mm}
	\left(
	\prod_{a=1}^{J}
	(p_{i_a}\cdot\lambda)\:
	c_{k_a}^{(i_a)}
	\right)
	\left(
	\prod_{b=1}^{J}
	(p_{j_b}\cdot\lambda')\:
	d_{l_b}^{(j_b)}
	\right)
	B(-\alpha(s)+k+l,-\alpha(t)).
\end{align}
Here the coefficients $c_k^{(i)}, d_l^{(j)}$ are defined by
\begin{align}
	&c_k^{(2)} = +c_k(\alpha_2+1,\alpha_1+1,\alpha_2'), \\
	&c_k^{(2')} = -c_{k-1}(\alpha_2,\alpha_1,\alpha_2'+1), \\
	&d_l^{(2)} = -c_{l-1}(\beta_2',\beta_1',\beta_2+1), \\
	&d_l^{(2')} = +c_l(\beta_2'+1,\beta_1'+1,\beta_2),
\end{align}
where
\begin{align}
	&c_k(\alpha_2,\alpha_1,\alpha_2') \notag\\
	&= \frac{ \Gamma(-\alpha_2+1)\Gamma(-\alpha_1+1) }{ \Gamma(-\alpha_2-\alpha_1+2) }
	\frac{ \Gamma(\alpha_2'+k) }{ \Gamma(\alpha_2') }
	\frac{ \Gamma(-\alpha_2+1+k) }{ \Gamma(-\alpha_2+1) }
	\frac{ \Gamma(-\alpha_2-\alpha_1+2) }{ \Gamma(-\alpha_2-\alpha_1+2+k) }
	\frac{1}{\Gamma(k+1)}, 
\end{align}
and
\begin{align}
    \label{eq:alpha_beta}
	\alpha_i = -(-N_aq) \cdot p_i, \quad
	\beta_j = -(-N_b'q') \cdot p_j.
\end{align}
The last factor $B(-\alpha(s)+k+l,-\alpha(t))$ in \eqref{eq:amp_hes_tachyon_s-shifted_main} is the Beta function
\begin{align}
	& B(x,y) = \frac{ \Gamma(x)\Gamma(y) }{ \Gamma(x+y) }, 
\end{align}
and $k$, $l$ and $\alpha$ are defined by 
\begin{align}
	\alpha(x) &= 1+x/2, &
	k &= \sum_{a=1}^{J} k_a, &
	l &= \sum_{b=1}^{J} l_b.
\end{align}

To evaluate the amplitude \eqref{eq:amp_hes_tachyon_s-shifted_main},
we have taken the following parameterization 
of the momenta and polarization vectors, 
\begin{align}
	\begin{array}{ll}
		-q = \frac{ 1 }{ \sqrt{2(N-1)+p^2}-p\cos\theta } \mqty(1\\0\\0\\-1), \qquad &
		-(-q') = \frac{ 1 }{ \sqrt{2(N-1)+p^2}-p\cos\theta' } \mqty(1\\0\\0\\-1), \\
		\lambda = \frac{1}{\sqrt{2}} \mqty(0\\1\\i\\0) &
		-\lambda' = \frac{1}{\sqrt{2}} \mqty(0\\1\\i\\0), \\
		p_1 = \mqty( \sqrt{p^2-2} \\ p\sin\theta \\ 0 \\ p\cos\theta ), &
		-p_1' = \mqty( \sqrt{p^2-2} \\ p\sin\theta'\cos\varphi' \\ p\sin\theta'\sin\varphi' \\ p\cos\theta' ), 	\\
		p_1+p_2-Nq = \mqty( \sqrt{s} \\ 0 \\ 0 \\ 0 ), &
		-p_1'-p_2'+N'q' = \mqty( \sqrt{s} \\ 0 \\ 0 \\ 0 ).
	\end{array}
\end{align}
Here, $p$ stands for the magnitude of the momentum of the incoming tachyon, 
and the scattering amplitude is parametrized by 
the angles $\theta$, $\theta'$ and $\varphi'$. 
The Mandelstam(-like) variables are now expressed as 
\begin{align}
	s &= (\sqrt{2(N-1)+p^2}+\sqrt{p^2-2})^2, \\
	t &= (\sqrt{2(N-1)+p^2}-\sqrt{p^2-2})^2
	- 2p^2 (1 + \cos\theta\cos\theta' + \sin\theta\sin\theta' \cos\varphi'), \\
	u &= - 2p^2 (1 - \cos\theta\cos\theta' - \sin\theta\sin\theta' \cos\varphi').
\end{align}
It is straightforward to check that these variables 
satisfy the identify \eqref{Mandel-id}. 
Also we easily find for \eqref{eq:alpha_beta} that
\begin{align}
    \label{eq:pq}
	\begin{array}{lll}
		p_1\cdot q = +\frac{ \sqrt{2(-1)+p^2}+p\cos\theta }{ \sqrt{2(N-1)+p^2}-p\cos\theta }, &
		p_2\cdot q = +1, &
		p_2'\cdot q = -\frac{ \sqrt{2(N-1)+p^2}-p\cos\theta' }{ \sqrt{2(N-1)+p^2}-p\cos\theta },\\
		%
		p_1' \cdot q' = +\frac{ \sqrt{2(-1)+p^2}+p\cos\theta' }{ \sqrt{2(N-1)+p^2}-p\cos\theta' }, &
		p_2'\cdot q' = +1, &
		p_2 \cdot q' = -\frac{ \sqrt{2(N-1)+p^2}-p\cos\theta }{ \sqrt{2(N-1)+p^2}-p\cos\theta' },
	\end{array}	
\end{align}
and for the coefficients in \eqref{eq:amp_hes_tachyon_s-shifted_main} that
\begin{align}
    &p_2\cdot\lambda
    = -p_2\cdot\lambda'
    = \frac{-1}{\sqrt{2}}\: p\sin\theta, \\
    &p_2'\cdot\lambda'
    = -p_2'\cdot\lambda
    = \frac{-1}{\sqrt{2}}\: p e^{i\varphi'} \sin\theta'.
\end{align}
Under these parametrization,
the HES-tachyon to HES-tachyon amplitude
is given by
the momentum and the incoming/outgoing angles
$\mathcal{A}=\mathcal{A}(\{N_a\},\{N_b'\};p,\theta,\theta',\varphi')$.

We apply the strategy in Sec.~\ref{sec:gen_quantum_scattering}
to this amplitude and find the largest pole, to extract the scattering data.
Let us consider the amplitude for given 
excitation levels $\{N_a\},\{N_b'\}$ and magnitude of momentum $p$.
Since the Mandelstam variable $s$ depends only 
on the total excitation level $N$ and magnitude of momentum $p$, 
the $s$-channel poles are independent of the angular variables. 
On the other hand, the $t$-channel poles come from 
the divergences of the Beta function in \eqref{eq:amp_hes_tachyon_s-shifted_main}, 
and appear in certain outgoing angles for given incoming angles. 
The $t$-channel poles form a line in two-dimensional space 
of outgoing angles $\theta'$ and $\varphi'$. 
For a fixed $\varphi'$, 
the positions of poles in the outgoing angle $\{\theta'_m(\theta)\}$ 
is given as functions of the incoming angle $\theta$. 
They can be obtained by solving
\begin{align}
    \label{eq:t-pole_eq}
    \alpha(t(\theta,\theta')) = n, \quad
    n \in \mathbb{Z}_{\geq}, 
\end{align}
as the $t$-channel poles come from the poles of $\Gamma(-\alpha(t))$ in 
the Beta function in \eqref{eq:amp_hes_tachyon_s-shifted_main}. 
We evaluate this equation numerically.
Among the data $\{\theta_m'(\theta)\}$,
we look for their maximum poles, which has the largest residue of the poles. 
The residue around the poles is
\begin{align}
    \tilde{\mathcal{A}}_{st}
    &=
    \lim_{ \alpha(t)\rightarrow n }
    \lim_{ \{\alpha_2\} \rightarrow \{N_a\} }
    \lim_{ \{\beta_2'\}\rightarrow \{N_b'\} }
    \frac{\sin\pi(-\alpha(t))}{\pi}
    \prod_{a=1}^{J}
    \frac{\sin\pi\alpha_2}{\pi}
    \prod_{b=1}^{J}
    \frac{\sin\pi\beta_2'}{\pi}\:
    \mathcal{A}_{st} \notag\\
    &=
    \left. \mathcal{A}_{st} \right|_{ B\rightarrow \tilde{B}, c\rightarrow\tilde{c}, d\rightarrow\tilde{d} }.
\end{align}
Here we have defined
\begin{align}
	\tilde{B}(-\alpha(s)+k+l,-\alpha(t))
	&= \lim_{\alpha(t)\rightarrow n}
	\frac{ \sin\pi(-\alpha(t)) }{\pi}\:
	B(-\alpha(s)+k+l,-\alpha(t)) \notag\\
	&=
	\frac{ \Gamma(-\alpha(s)+k+l) }
	{ \Gamma(-\alpha(s)-n+k+l)\Gamma(1+n) }, \\
	\tilde{c}_k(\alpha_2,\alpha_1,\alpha_2')
	&= \lim_{\alpha_2 \rightarrow N_a}
	\frac{ \sin\pi\alpha_2 }{ \pi }\:
	c_k(\alpha_2,\alpha_1,\alpha_2') \notag\\
	&=
	\frac{1}{\Gamma(k+1)\Gamma(\alpha_2-k)}
	\frac{\Gamma(\alpha_2-k+\alpha_1-1)}{\Gamma(\alpha_1)}
	\frac{\Gamma(\alpha_2'+k)}{\Gamma(\alpha_2')},
\end{align}
and have also defined
\begin{align}
	&\tilde{c}_k^{(2)} = -\tilde{c}_k(\alpha_2+1,\alpha_1+1,\alpha_2'), \\
	&\tilde{c}_k^{(2')} = -\tilde{c}_{k-1}(\alpha_2,\alpha_1,\alpha_2'+1), \\
	&\tilde{d}_l^{(2)} = -\tilde{c}_{l-1}(\beta_2',\beta_1',\beta_2+1), \\
	&\tilde{d}_l^{(2')} = -\tilde{c}_l(\beta_2'+1,\beta_1'+1,\beta_2).
\end{align}
Note that the signs of the coefficients $\tilde{c}^{(2)}, \tilde{d}^{(2')}$ are flipped
since $\sin\pi(\alpha_2+1)=-\sin\pi\alpha_2$.
Numerically evaluating the reside
at $\{\theta'_m(\theta)\}$,
and finding the largest one among them,
we construct the scattering data $\theta'(\theta)$
of the HES-tachyon scattering amplitudes.

\subsection{Transient chaos analysis of the scattering amplitude}

In this subsection,
we show
numerical plots of the scattering data $\theta'(\theta)$
for various fixed parameters,
and discuss their fractal feature (the self-similarity).

\begin{figure}[t]
	\centering
	\includegraphics[width=65mm]{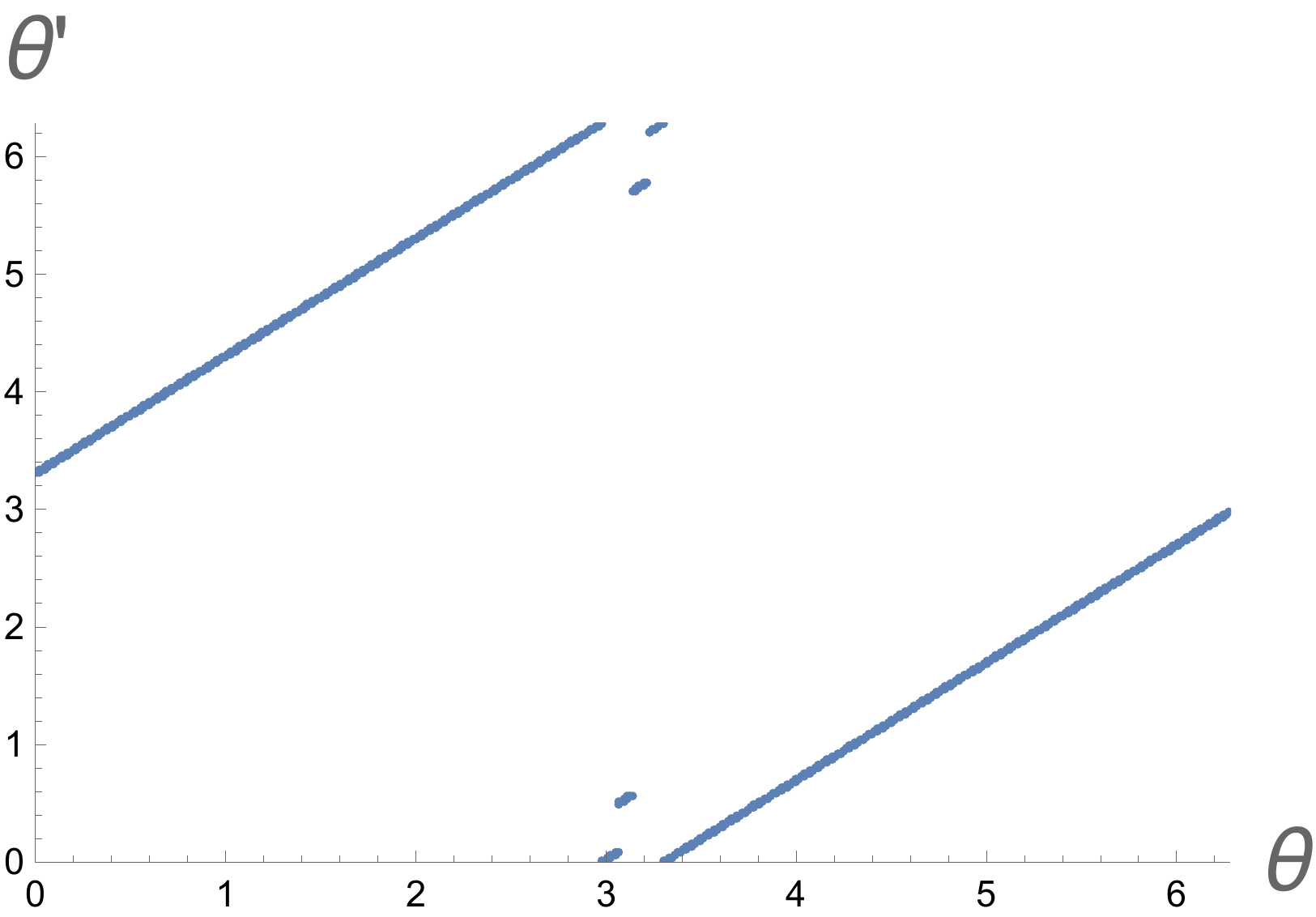} \hspace{5mm}
	\includegraphics[width=65mm]{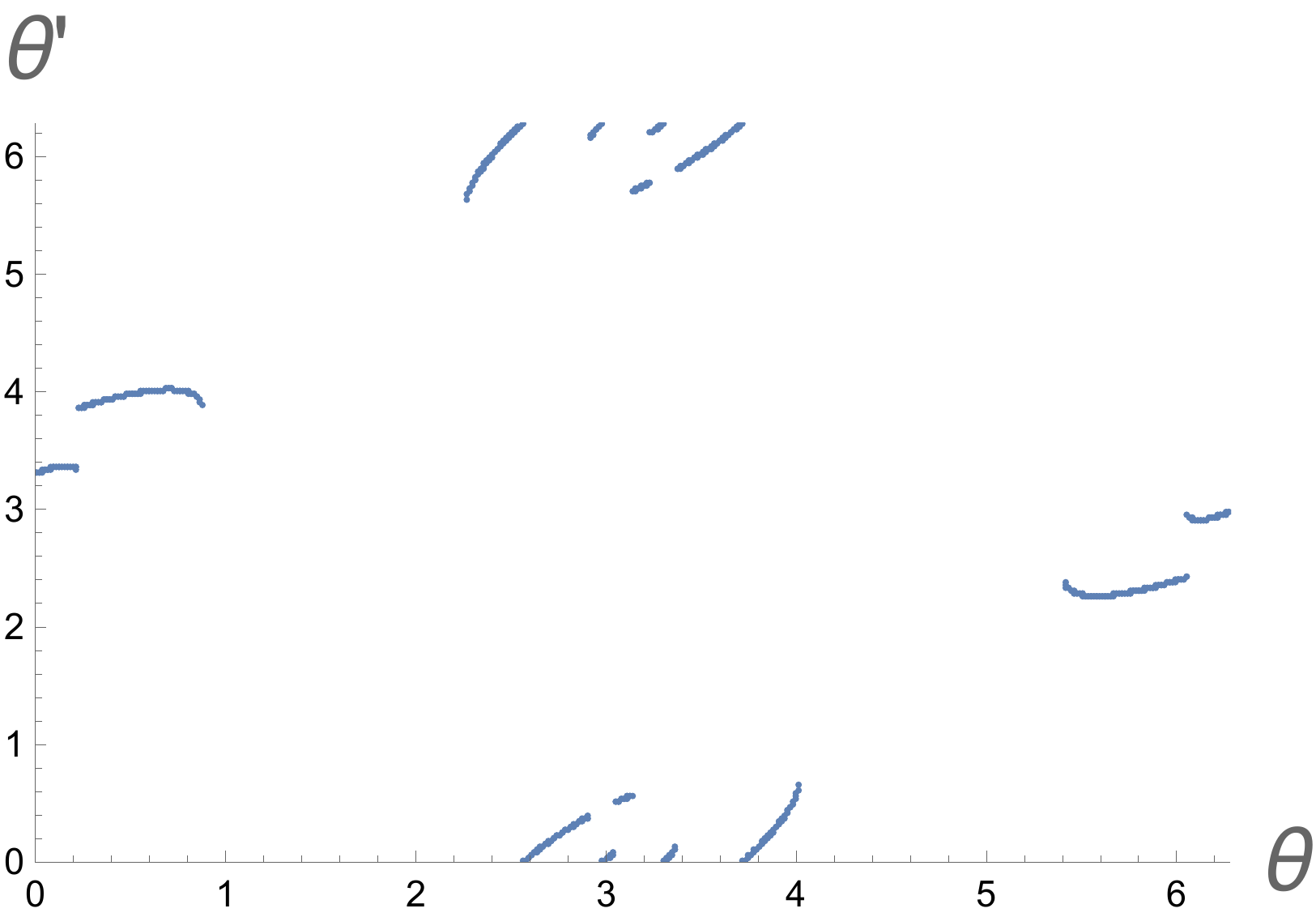} \\ \vspace{3mm}
	\includegraphics[width=65mm]{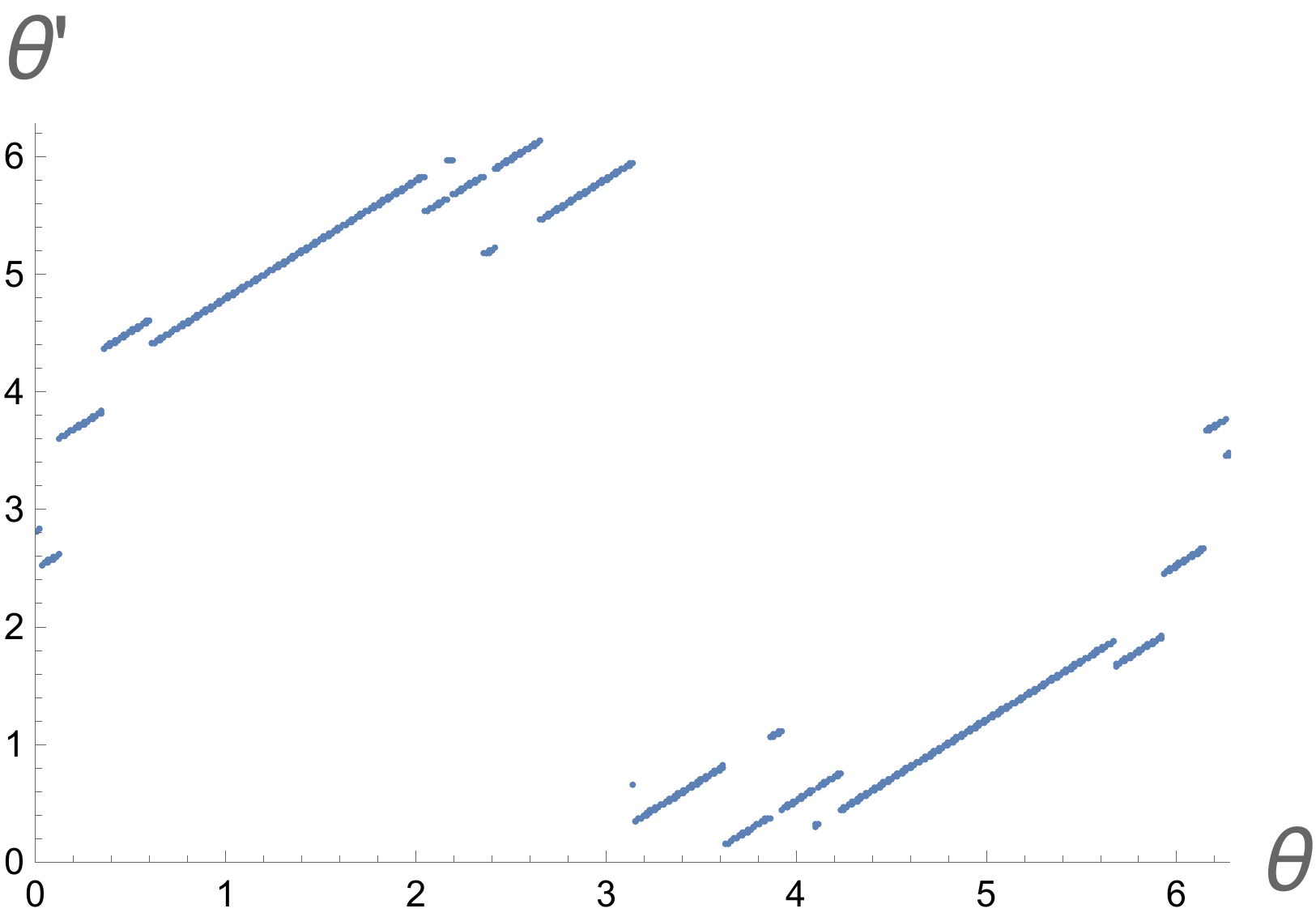} \hspace{5mm}
	\includegraphics[width=65mm]{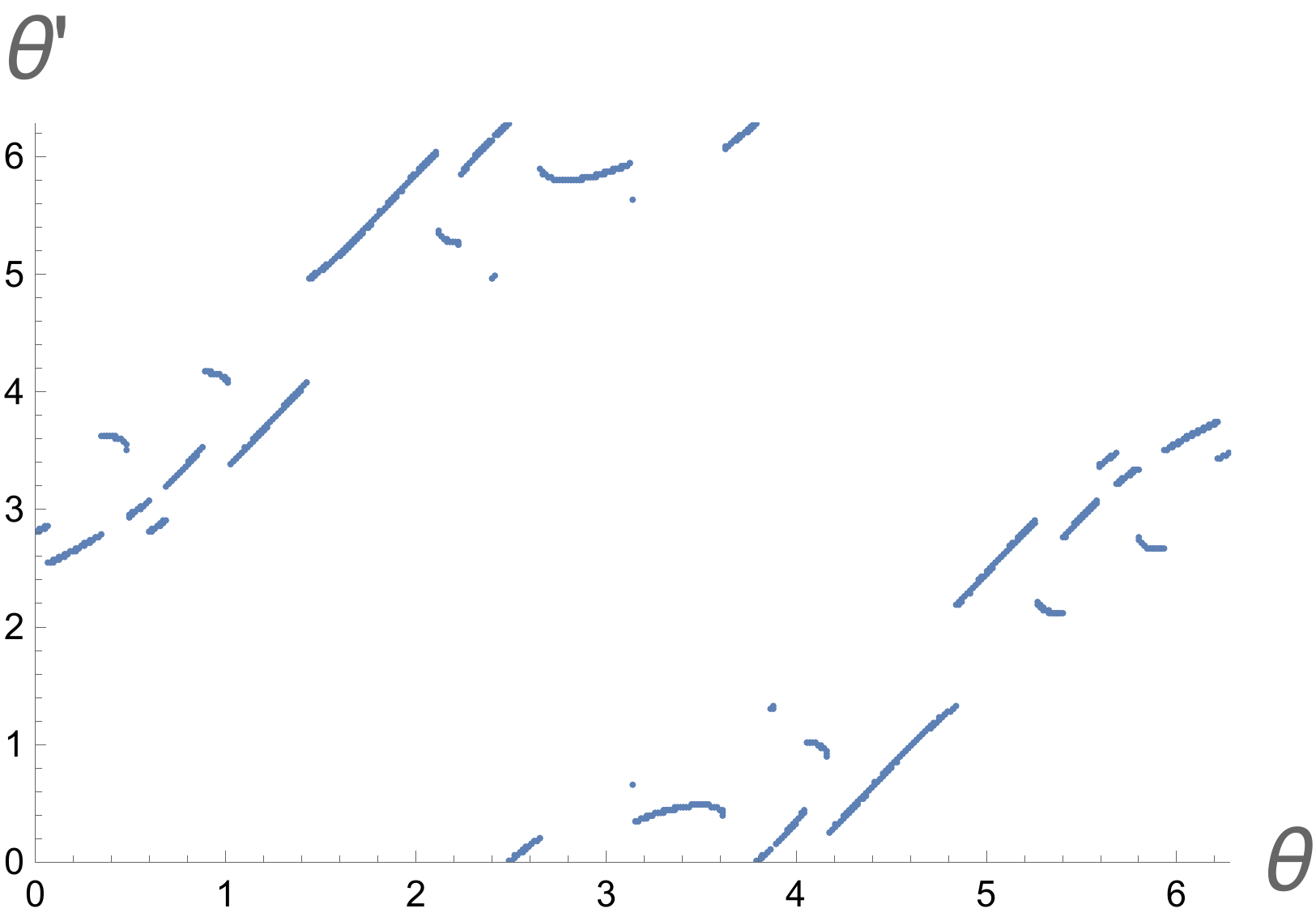} \\ \vspace{3mm}
	\includegraphics[width=65mm]{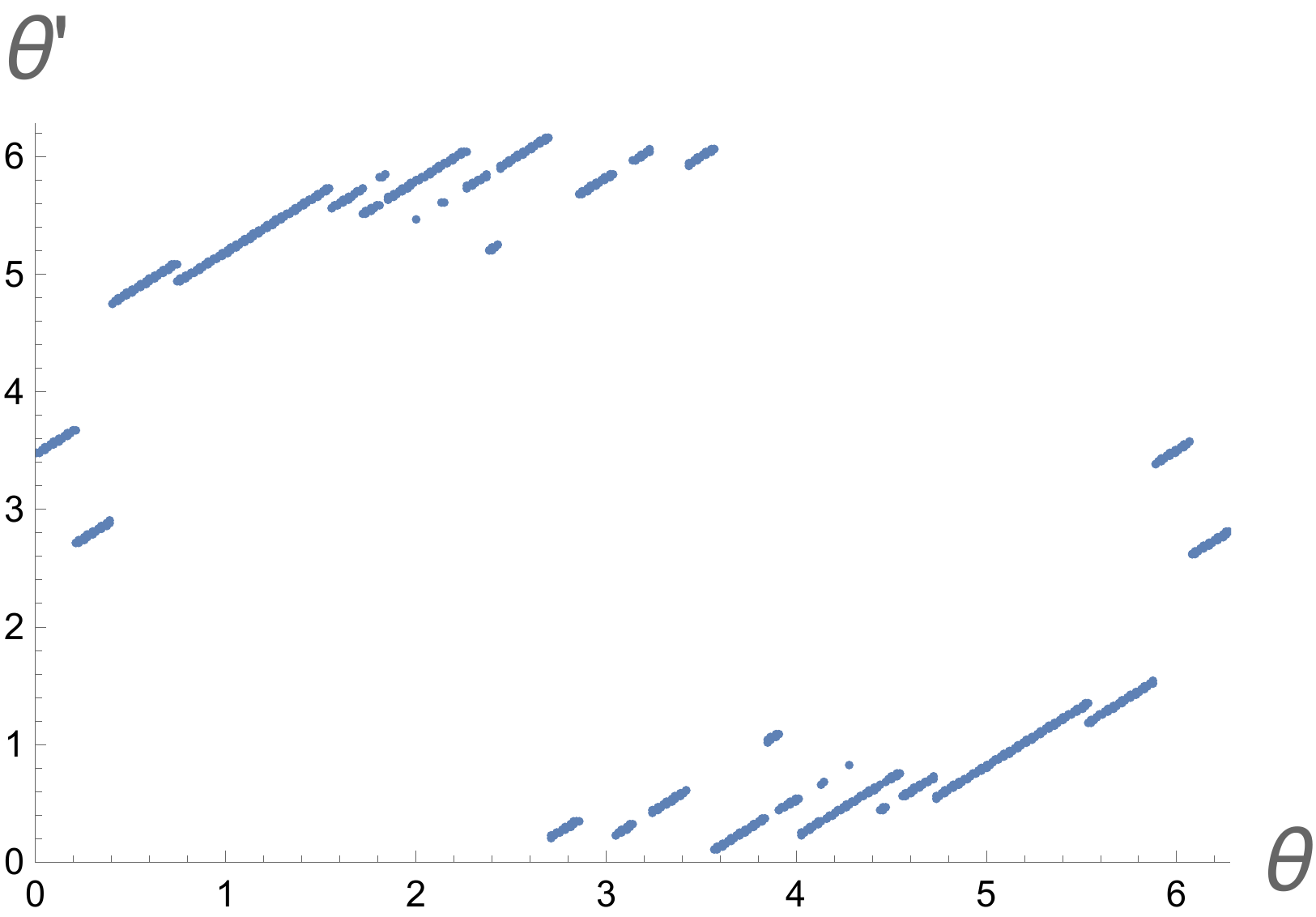} \hspace{5mm}
	\includegraphics[width=65mm]{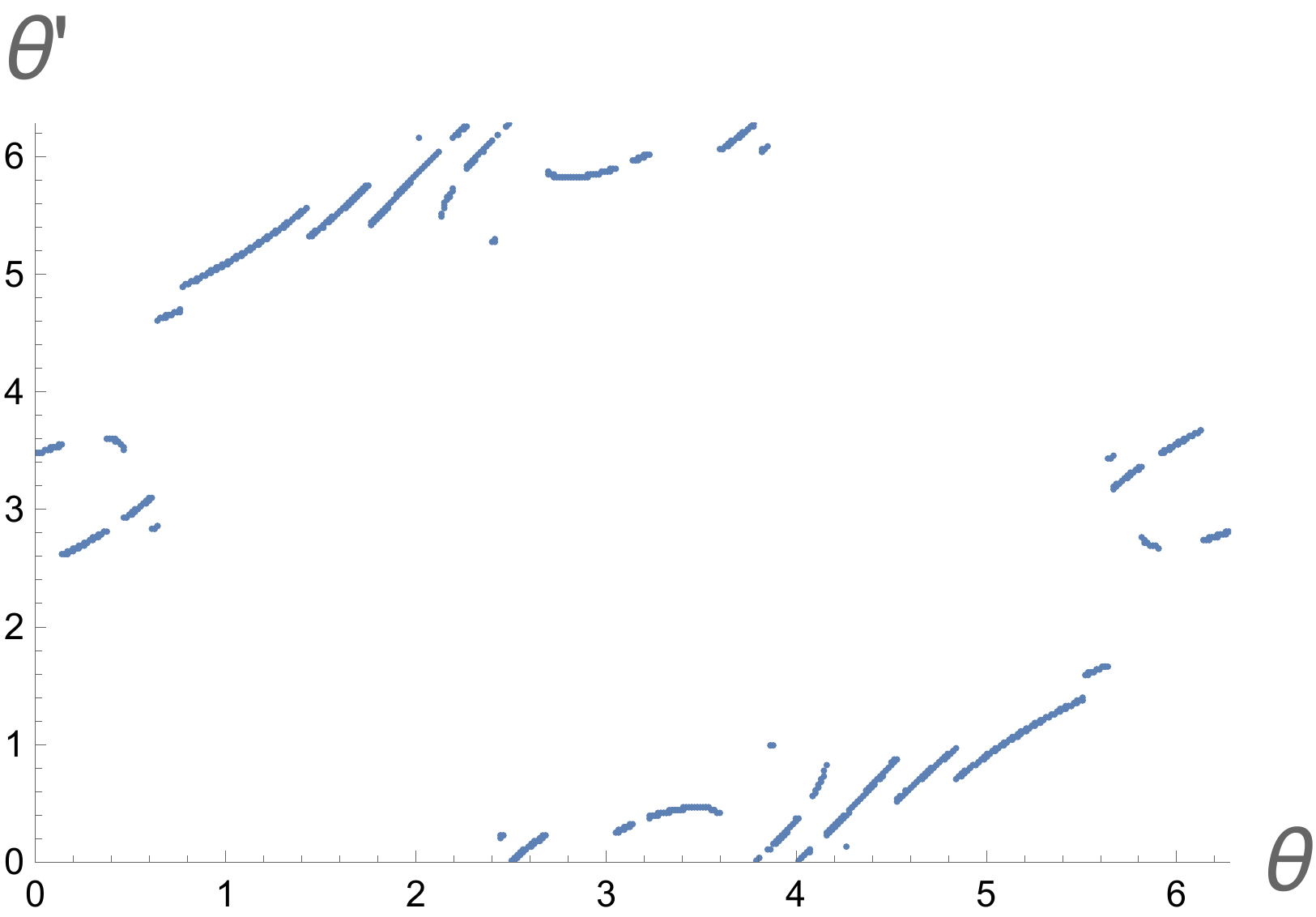} \\ \vspace{3mm}
	\caption{
	    Scattering data $\theta'(\theta)$
	    involving a HES with the angular momentum $J=1$.
	    The top/middle/bottom two panels
	    are the plots when the excitation level is $N=N_1=N_1'=1,5,9$.
	    The three panels on the left/right hand side
	    are the plots when $\varphi'=0,\pi/4$.
	    The momentum is fixed
	    at $p=2.6$.
	    Although the plots are
	    highly erratic,
	    fractal structure has not been observed.
	}
	\label{fig:theta_thetap_J=1}
\end{figure}

\begin{figure}[t]
	\centering
	\includegraphics[width=65mm]{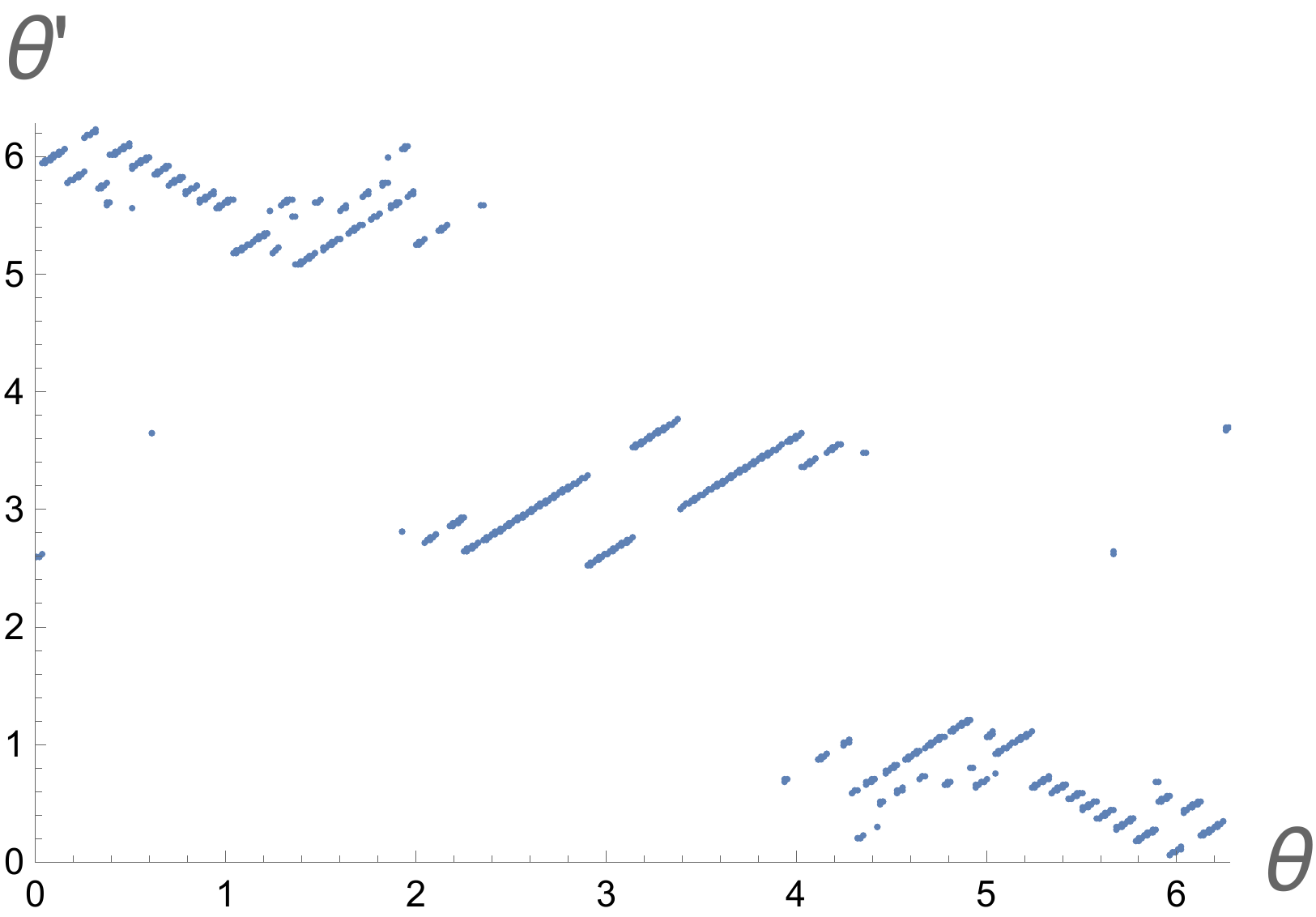} \hspace{5mm}
	\includegraphics[width=65mm]{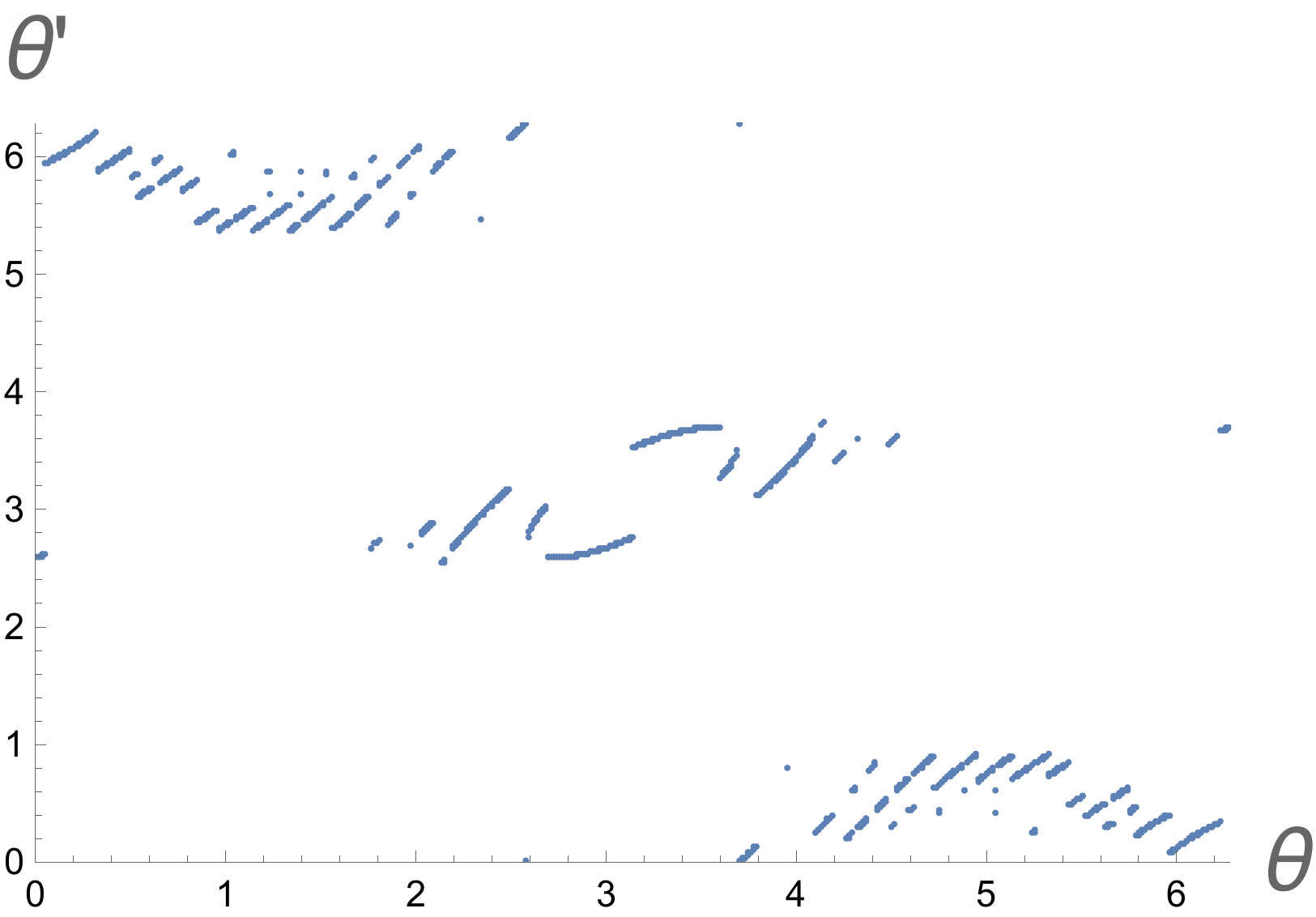}
	\caption{
	    Scattering data $\theta'(\theta)$
	    when $J=1,N=27$.
	    The excitation level
	    is chosen so that as many $t$-channel poles appear as possible.
	    Other parameters are fixed as Fig.~\ref{fig:theta_thetap_J=1}.
	    Still fractal structure does not appear. 
	}
	\label{fig:theta_thetap_J=1_higher}
\end{figure}

The results for a scattering
of a tachyon and a HES with the angular momentum $J=1$
are shown in Fig.~\ref{fig:theta_thetap_J=1}.
The top/middle/bottom two panels
are the plots
for the excitation levels $N=N_1=N_1'=1,5,9$.
The three panels on the left and right hand side
are the plots for $\varphi'=0$ and $\varphi = \pi/4$, respectively.
The momentum is
fixed at $p=2.6$
so that the energy squared is positive,
$E^2=p^2-2 \approx 7-2 > 0$.

The plots
are highly erratic,
for larger excitation levels.
The tilted lines
in the case $N=1$
split into smaller segments
in the cases $N=5,9$.
However,
we do not see any fractal structure.
In particular, the number of the small segments
does not increase exponentially in $N$. Note that typical fractal data is produced by a repeated action of an operation creating self-similar structure, and the number of small structure grows exponentially in the number of the action, thus also in our case, it is expected that the number of the smaller segments would have grown exponentially in $N$ if the scattering were chaotic.

This is further confirmed  
in Fig.~\ref{fig:theta_thetap_J=1_higher}, where the excitation level
is chosen as $N=27$. This number $N=27\approx 4p^2$ was chosen
so that as many $t$-channel poles,
found in \eqref{eq:t-pole_eq},
appear as possible, 
nevertheless, 
the fractal feature, or the exponentially smaller structure,
does not appear.
Thus
we conclude that
the scattering data is not fractal.
This means that our HES-tachyon scattering is not chaotic.

\begin{figure}[t]
	\centering
	\includegraphics[width=65mm]{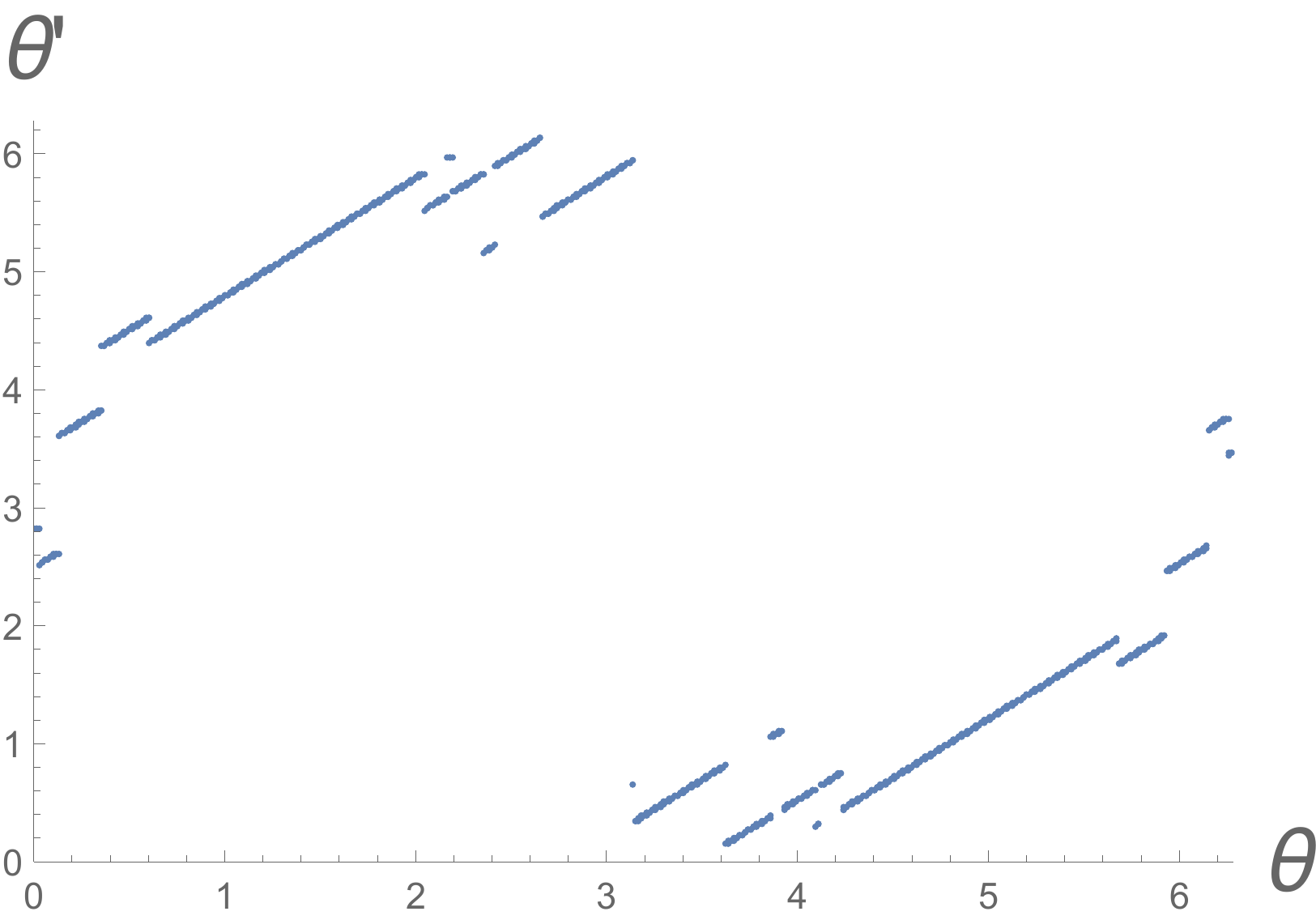} \\
	\includegraphics[width=65mm]{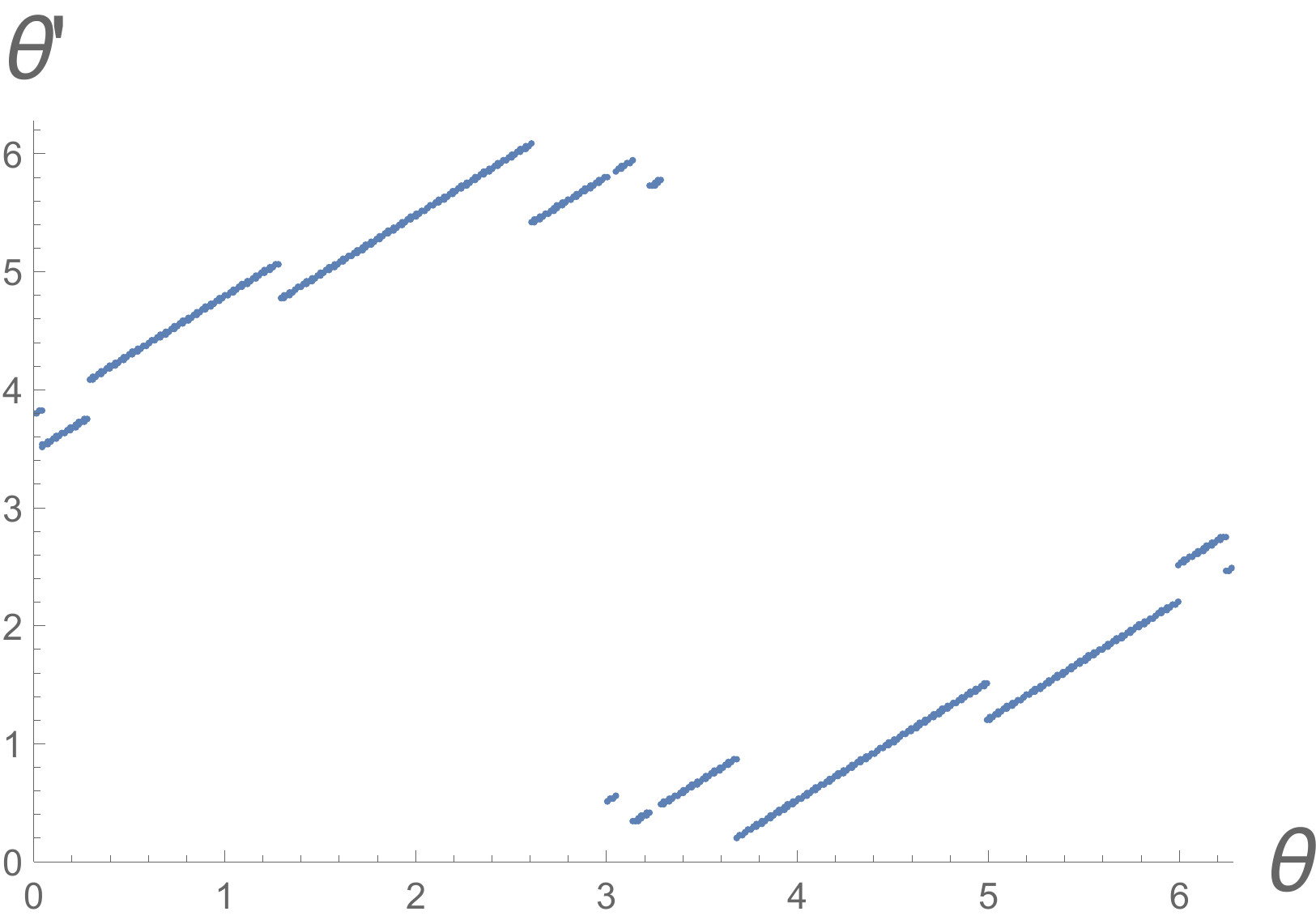}
	\includegraphics[width=65mm]{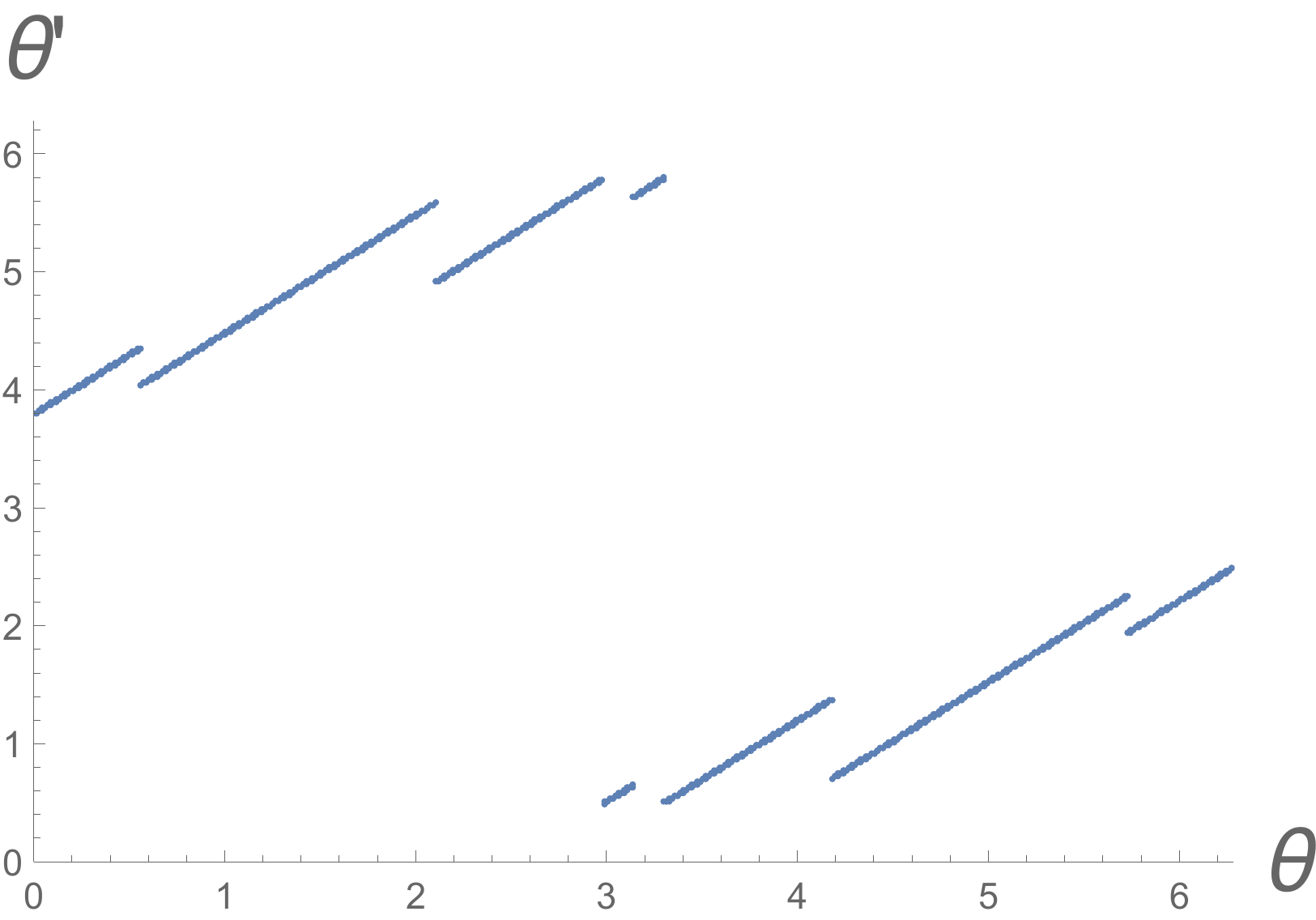}
	\caption{
	    Scattering data $\theta'(\theta)$
	    involving a HES with the angular momentum $J>1$.
	    The top/bottom-left/bottom-right panels
	    are the plots
	    when the excitation level is
	    $\{N_a\}=\{N_b'\}=\{5\},\{3,1,1\},\{1,1,1,1,1\}$.
	    Other parameters are fixed as
		$\varphi'=0, p=2.6$.
		As the level $N$
		is partitioned into smaller numbers $\{N_a\},\{N_b'\}$,
		the small segments disappear.
	}
	\label{fig:theta_thetap_J=gen}
\end{figure}

The results for the scattering of the HES with $J>1$
are shown in Fig.~\ref{fig:theta_thetap_J=gen}.
The top, bottom-left and bottom-right panels
are the plots for the excitation levels $\{N_a\}=\{N_b'\}=\{5\},\allowbreak\{3,1,1\},\{1,1,1,1,1\}$, respectively.
Other parameters are fixed as
$\varphi'=0$, $p=2.6$.
As the photons of which the HES with the total excitation level $N$ consists
are partitioned into more photons to form a HES  
with smaller excitation levels $\{N_a\},\{N_b'\}$,
we find in Fig.~\ref{fig:theta_thetap_J=gen} that the small segments disappear.
We conclude that even with the HES with $J>1$ the fractal nature is not found,
therefore, the scattering amplitudes are not chaotic.

This behavior on the partitions is explained as follows.
As seen from the examples of the case $J=1$, 
the smaller segments appear for larger $N$ (or $N_a, N_b'$). 
In the scattering amplitude, there exist several lines of poles. 
When we pick out the pole with the largest residue, 
the largest pole depends on the incoming angles. 
Thus, the line in the scattering data of the largest pole is segmented. 
This magnitude of the residue at each pole is mostly determined by 
the coefficients $c_k^{(i)}(\alpha), d_l^{(j)}(\beta)$,
where $\alpha,\beta$ are proportional to
$N_a, N_b'$ times oscillating functions
(see \eqref{eq:alpha_beta}, \eqref{eq:pq}). 
These coefficients oscillate more rapidly
for larger values of $\{N_a\},\{N_b'\}$.
Then, the largest pole frequently changes from 
a line to another, and then, the line of 
the largest pole is split into smaller segments. 
The minimum length of the segments is bounded by 
the maximum of $\{N_a\},\{N_b'\}$, 
and hence, cannot be exponentially small. 
This is the reason why the scattering data is not fractal.

\section{Conclusion and discussions}
\label{sec:conclusion}

In this paper, following the observation of the erratic nature of the decay amplitudes
of highly excited strings \cite{Gross:2021gsj,Rosenhaus:2021xhm}, 
we have proposed a way to define the scattering data of 
quantum scattering amplitudes in open bosonic string theory, and have looked for
any fractal structure in the scattering data of the scattering of
a tachyon and a highly excited string (HES). Our search has gone up to the HES excitation
level $N=27$ with various manners of forming the HES states. 
The scattering data of the tachyon-HES amplitudes is summarized 
in Fig.~\ref{fig:theta_thetap_J=1}, 
Fig.~\ref{fig:theta_thetap_J=1_higher} and Fig.~\ref{fig:theta_thetap_J=gen}.
Within our search, we could not identify any fractal structure in the scattering data.
Since the fractal structure in the scattering data is one of the established diagnoses 
for pinning down the existence of transient chaos, we have come to the observation 
that the string scattering is not chaotic, within our search and our definition.

Nevertheless, it is too early to admit that string scattering has no chaos.
Our HES states are made by following the construction given in \cite{Gross:2021gsj}, while
generic HES can have more degrees of freedom. For example, the photons which are introduced 
to create the HES have shared the identical momentum/polarization in our HES, 
and they could be generalized. In fact, as shown in \cite{Hashimoto:2022ugt}, the
imaging of the HES by the decay amplitude of the HES to 2 tachyons, with the same definition
of the HES states, results in just a set of slits which are aligned on a line in the target space.
Although the classical scattering by the four-hill potential 
in Sec.~\ref{sec:review} shows chaos because 
the particle is scattered many times by the potential, 
such multiple scattering would not happen if the hills were aligned on a line. 
The imaging of the HES decay amplitude implies that 
it is difficult for any tachyon probe to be scattered many times by the HES.
The HES decay amplitudes should have more complicated, 
namely, higher dimensional structures to have chaos. 
The multiple scattering is the essence of the transient chaos, thus the imaging results of
the HES in \cite{Hashimoto:2022ugt} suggests that more general HES states are needed to 
produce any transient chaos. In other words, if the HES is the one formed by more variety
of strings, the HES can be complicated enough to produce the chaos. This also translates to
the consideration of 5-point or 6-point amplitudes.

One of the other possibilities to generate chaos is the modification of 
the definition of the scattering data.
Since our definition of the scattering data extracts just an aspect of the quantum amplitudes (although we believe that it is a natural definition), there could be
more sophisticated definition of the scattering data which may manifest possible
transient chaos. For example, we have ignored the impact parameter $b$ in our definition
because usually the quantum scatterings are defined by plane waves. However, if
one considers wave packets rather than the plane waves, the amplitudes would 
contain more information about the locality of the HES structure.

If we regard the multiple scattering in classical transient chaos as multiple scattering in
string theory, it could mean that we need to visit higher loop amplitudes.
The scattering amplitudes which we have studied in this paper is the tree level scattering,
which is of order of the string coupling constant. However, the particle scattering by a fixed potential is a non-perturbative process in the sense that the scattering potential can be decomposed into its radial modes. Therefore, naively speaking, for a possible transient chaos, 
higher loop amplitudes would be necessary. 
As a first guess, we can use the tree level scattering data given 
in Fig.~\ref{fig:theta_thetap_J=1}, 
Fig.~\ref{fig:theta_thetap_J=1_higher} and Fig.~\ref{fig:theta_thetap_J=gen}
to produce the multiple scattering. For example, we can use the function $\theta'(\theta)$
and apply this map many times, as $\theta'(\theta'(\theta)), \cdots$. This is an analogy of
the popular Baker's map which produces the chaos.
Unfortunately, since the slope of the scattering data in the left panels of Fig.~\ref{fig:theta_thetap_J=1} and Fig.~\ref{fig:theta_thetap_J=1_higher} and in 
Fig.~\ref{fig:theta_thetap_J=gen} is identical to 1, this map does not work to magnify a part
of the scattering data (which is necessary for any chaotic map). Therefore, at this stage,
our naive multiple application of the scattering data map does not seem to produce the chaos.\footnote{The slope of some parts of the scattering data with $\varphi'=\pi/4$ exceeds 1,
and the measure analysis is necessary to figure out whether the map can generate chaos or not.}

To explicitly formulate the correspondence principle between a black hole and highly excited strings, we need to pin down the place where the chaos shows up in string perturbation theory. 
One may need some statistical approach. Once the transient chaos is spotted, one may proceed to  
derive the Lyapunov exponent to be compared with the surface gravity of the black holes.\footnote{Recent result \cite{Bianchi:2022mhs} relates the HES decay amplitudes and the Wigner-Dyson statistical spectrum which is typical for quantum chaos. Any possible relation
to the fractal feature would be of importance.}
The quest for the formulation of the correspondence principle continues, and we hope that our 
present work serves as a step toward the formulation of it.

\subsection*{Acknowledgment}

The work of K.~H.\ was supported in part by JSPS KAKENHI Grant No.~JP22H01217, JP22H05111 and JP22H05115.
The work of Y.~M.\ was supported in part by JSPS KAKENHI Grant No.~JP20K03930, JP21H05182 and JP21H05186. 
The work of T.~Y.\ was supported in part by JSPS KAKENHI Grant No.~JP22J15276.
The work of K.~H.\ and Y.~M.\ was also supported in part by JSPS KAKENHI Grant No.~JP17H06462.
K.~H.\ would like to thank Yannick Paget for discussions and collaborative artwork concerning this research.

\appendix

\section{Explicit formulas of HES-tachyon to HES-tachyon amplitude}
\label{sec:hes-tachyon_amp}

We start with a general formula
for the bosonic open string tree-level amplitude,
and pick out poles
corresponding to HES states
as was done in \cite{Rosenhaus:2021xhm}.
We use the same strategy
to compute the four-point amplitude of the scattering of 
the HES-tachyon to HES-tachyon, 
to obtain the scattering data $\theta'(\theta)$.

\subsection{Channels}

The general formula
for the tree-level amplitude
involving any number of tachyons and photons is
given by
\begin{align}
	\label{eq:gen_amp}
	\mathcal{A}
	&= \frac{1}{\text{vol.}}
	\int \dd{w_i}\dd{z_a}
	\prod_{i<j} \abs{w_{ij}}^{p_i\cdot p_j}
	\prod_{a<b} \abs{z_{ab}}^{p_a\cdot p_b}
	\prod_{i,a} \abs{w_i-z_a}^{p_i\cdot p_a} \notag\\
	&\hspace{60mm}\times
	\left.\exp\left[
	\sum_{a\neq b}\frac{\lambda_a\cdot\lambda_b}{2z_{ab}^2}
	-\sum_{i,a} \frac{p_i\cdot\lambda_a}{w_i-z_a}
	\right] \right|_{\mathcal{O}(\lambda_a)},
\end{align}
where $w_{ij}=w_i-w_j,\: z_{ab}=z_a-z_b$.
Here the indices $i$ run over the tachyons,
and the indices $a$ run over the photons.
The photon polarizations are denoted by $\lambda_a$.

We focus on the case $q\propto q'$ 
but applied to general integer $J$.\footnote{
See also \cite{Bianchi:2019ywd} for a recent study on computation of the amplitudes.}
Using the assumption $q\propto q'$,
the amplitude reduces to
\begin{align}
	\label{eq:hes_amp}
	\mathcal{A}
	&= \frac{1}{\text{vol.}}
	\int_{-\infty}^{\infty} \prod_{i=1,2,2',1'} \dd{w_i}\;
	\prod_{i<j} \abs{ w_{ij} }^{ p_i\cdot p_j } \notag\\
	&\hspace{30mm}\times
	\int_{-\infty}^{\infty} \prod_{a=1}^{J} \dd{z_a}\;
	\prod_{ a,i } \abs{ z_a-w_i }^{-\alpha_i}
	\sum_{ i }\frac{ -p_i\cdot\lambda }{ w_i-z_a } \notag\\
	&\hspace{60mm}\times
	\int_{-\infty}^{\infty} \prod_{b=1}^{J} \dd{z_b'}\;
	\prod_{ b,j } \abs{ z_b'-w_j }^{-\beta_j}
	\sum_{ j }\frac{ -p_j\cdot\lambda' }{ w_j-z_b' }
\end{align}
where we defined
\begin{align}
	\alpha_i = -(-N_aq) \cdot p_i, \quad
	\beta_j = -(-N_b'q') \cdot p_j.
\end{align}
The integral over $w_i$
can be divided into six parts
as in the case of the Veneziano amplitude,
\begin{align}
	\mathcal{A}
	&=
	\left( \mathcal{A}_{st}+\mathcal{A}_{tu}+\mathcal{A}_{us} \right)
	+\left( \mathcal{A}_{ts}+\mathcal{A}_{ut}+\mathcal{A}_{su} \right),
\end{align}
where
\begin{align}
	\label{eq:hes_amp_st}
	\mathcal{A}_{st}
	&= \left. \mathcal{A} \right|_{w_1'=-\infty,w_2=0,w_1=w,w_2'=1} \notag\\
	&= \int_{0}^{1} \dd{w}\;
	w^{p_1\cdot p_2} (1-w)^{p_1\cdot p_2'}\:
	\prod_{a=1}^{J}
	Z_a^{212'}(\alpha, p, \lambda; w)
	\prod_{b=1}^{J}
	Z_b^{212'}(\beta, p, \lambda'; w) \\
	\mathcal{A}_{tu}
	&= \left. \mathcal{A} \right|_{w_2=-\infty,w_2'=0,w_1=w,w_1'=1}
	= \left. \mathcal{A}_{st} \right|_{ 2\rightarrow 2', 2' \rightarrow 1' }, \\
	\mathcal{A}_{us}
	&= \left. \mathcal{A} \right|_{w_2'=-\infty,w_1'=0,w_1=w,w_2=1}
	= \left. \mathcal{A}_{st} \right|_{ 2\rightarrow 1', 2' \rightarrow 2 }, \\
	\mathcal{A}_{ts}
	&= \left. \mathcal{A}_{st} \right|_{ 2 \leftrightarrow 2' }, \\
	\mathcal{A}_{ut}
	&= \left. \mathcal{A}_{tu} \right|_{ 2' \leftrightarrow 1' }, \\
	\mathcal{A}_{su}
	&= \left. \mathcal{A}_{us} \right|_{ 1' \leftrightarrow 2 }.
\end{align}
Here we have defined the photon integral
\begin{align}
	Z_a^{ijk}(\alpha,p,\lambda; w)
	&=
	\int_{-\infty}^{\infty} \dd{z_a}\;
	\abs{z_a}^{-\alpha_i} \abs{z_a-w}^{-\alpha_j} \abs{z_a-1}^{-\alpha_k}
	\left(
	\frac{ -p_i\cdot\lambda }{ -z_a }
	+\frac{ -p_j\cdot\lambda }{ w-z_a }
	+\frac{ -p_k\cdot\lambda }{ 1-z_a }
	\right)
\end{align}

\subsection{Photon integral}

To proceed,
we define following integrals
and evaluate them.
\begin{align}
	\input{fig/st_s.tikz}
	&= \int_{0}^{w} \dd{z_a}\:
	\abs{z_a}^{-\alpha_2}\abs{z_a-w}^{-\alpha_1}\abs{z_a-1}^{-\alpha_2'} \notag\\
	&= \int_0^1 \dd{x}\:
	w^{1-\alpha_2-\alpha_1}
	\frac{ x^{-\alpha_2}(1-x)^{-\alpha_1} }{ (1-wx)^{\alpha_2'} }
	\quad (z=wx) \notag\\
	&= w^{1-\alpha_2-\alpha_1}
	\Gamma(-\alpha_2+1)\Gamma(-\alpha_1+1)\:
	\Freg{\alpha_2'}{-\alpha_2+1}{-\alpha_2-\alpha_1+2}{w}, \\
	\input{fig/st_tp.tikz}
	&= \int_{-\infty}^{0} \dd{z_a}\:
	\abs{z_a}^{-\alpha_2}\abs{z_a-w}^{-\alpha_1}\abs{z_a-1}^{-\alpha_2'} \notag\\
	&= \int_0^1 \dd{x}\:
	\frac{ x^{-2+\alpha_2+\alpha_1+\alpha_2'}(1-x)^{-\alpha_2} }{ (1-(1-w)x)^{\alpha_1} }
	\quad (1-z=1/x) \notag\\
	&= \Gamma(\alpha_2+\alpha_1+\alpha_2'-1)
	\Gamma(-\alpha_2+1)\:
	\Freg{\alpha_1}{\alpha_2+\alpha_1+\alpha_2'-1}{\alpha_1+\alpha_2'}{1-w}, \\
	\input{fig/st_t.tikz}
	&=\int_{w}^{1} \dd{z_a}\:
	\abs{z_a}^{-\alpha_2}\abs{z_a-w}^{-\alpha_1}\abs{z_a-1}^{-\alpha_2'} \notag\\
	&= \int_0^1 \dd{x}\:
	(1-w)^{1-\alpha_2'-\alpha_1}
	\frac{ x^{-\alpha_2'}(1-x)^{-\alpha_1} }{ (1-(1-w)x)^{\alpha_2} }
	\quad (z=(1-w)(1-x)+w) \notag\\
	&= (1-w)^{1-\alpha_2'-\alpha_1}
	\Gamma(-\alpha_2'+1)\Gamma(-\alpha_1+1)\:
	\Freg{\alpha_2}{-\alpha_2'+1}{-\alpha_2'-\alpha_1+2}{1-w} \notag\\
	&= \left. \input{fig/st_s.tikz} \right|_{ 2 \leftrightarrow 2',\: w \leftrightarrow 1-w }, \\
	\input{fig/st_sp.tikz}
	&= \int_{1}^{\infty} \dd{z_a}\:
	\abs{z_a}^{-\alpha_2}\abs{z_a-w}^{-\alpha_1}\abs{z_a-1}^{-\alpha_2'} \notag\\
	&= \int_0^1 \dd{x}\:
	\frac{ x^{-2+\alpha_2'+\alpha_1+\alpha_2}(1-x)^{-\alpha_2'} }{ (1-wx)^{-\alpha_1} }
	\quad (z=1/x) \notag\\
	&= \Gamma(\alpha_2'+\alpha_1+\alpha_2-1)
	\Gamma(-\alpha_2'+1)\:
	\Freg{\alpha_1}{\alpha_2'+\alpha_1+\alpha_2-1}{\alpha_1+\alpha_2}{w} \notag\\
	&= \left. \input{fig/st_tp.tikz} \right|_{ 2 \leftrightarrow 2',\: w \leftrightarrow 1-w }.	
\end{align}
Then the photon integrals are evaluated as follows.
\begin{align}
	\input{fig/st_s-s.tikz}
	&= \int_{0}^{w} \dd{z_a}\:
	\abs{z_a}^{-\alpha_2}\abs{z_a-w}^{-\alpha_1}\abs{z_a-1}^{-\alpha_2'}\:
	\frac{-p_2\cdot\lambda}{-z_a} \notag\\
	&= -(-p_2\cdot\lambda)
	\left. \input{fig/st_s.tikz} \right|_{\alpha_2 \rightarrow \alpha_2+1}, \\
	\input{fig/st_s-n.tikz}
	&= \int_{0}^{w} \dd{z_a}\:
	\abs{z_a}^{-\alpha_2}\abs{z_a-w}^{-\alpha_1}\abs{z_a-1}^{-\alpha_2'}\:
	\frac{-p_1\cdot\lambda}{w-z_a} \notag\\
	&= +(-p_1\cdot\lambda)
	\left. \input{fig/st_s.tikz} \right|_{\alpha_1 \rightarrow \alpha_1+1}, \\
	\input{fig/st_s-t.tikz}
	&= \int_{0}^{w} \dd{z_a}\:
	\abs{z_a}^{-\alpha_2}\abs{z_a-w}^{-\alpha_1}\abs{z_a-1}^{-\alpha_2'}\:
	\frac{-p_2'\cdot\lambda}{1-z_a} \notag\\
	&= +(-p_2'\cdot\lambda)
	\left. \input{fig/st_s.tikz} \right|_{\alpha_2' \rightarrow \alpha_2'+1}.
\end{align}
Other integrals over the intervals $(-\infty,0),(w,1),(1,\infty)$
are evaluated in a similar manner.

We are interested in the HES-tachyon to HES-tachyon scattering.
Picking out the HES-pole:
\begin{align}
	v,\: v' \sim 2(N-1),
	\quad\textit{i.e.}\quad
	\alpha_2 \sim N_a,\:
	\beta_2' \sim N_b',
\end{align}
the photon integral reduces to
\begin{align}
	&Z_a^{212'}(\alpha,p,\lambda;w)
	=\int_{-\infty}^{\infty} \dd{z_a}\;
	\abs{z_a}^{-\alpha_2} \abs{z_a-w}^{-\alpha_1} \abs{z_a-1}^{-\alpha_2'}
	\left(
	\frac{ -p_2\cdot\lambda }{ -z_a }
	+\frac{ -p_1\cdot\lambda }{ w-z_a }
	+\frac{ -p_2'\cdot\lambda }{ 1-z_a }
	\right) \notag\\
	&\sim
	\left(
	\input{fig/st_s-s.tikz}+
	\input{fig/st_tp-s.tikz}
	\right) \notag\\
	&\hspace{30mm}+
	\left(
	\input{fig/st_s-n.tikz}+
	\input{fig/st_tp-n.tikz}
	\right) \notag\\
	&\hspace{60mm}+
	\left(
	\input{fig/st_s-t.tikz}+
	\input{fig/st_tp-t.tikz}
	\right) \notag\\
	&=
	+(-p_2\cdot\lambda)
	\left.
	\left( 
	-\input{fig/st_s.tikz}
	+\input{fig/st_tp.tikz}
	\right)
	\right|_{\alpha_2\rightarrow\alpha_2+1} \notag\\
	&\hspace{30mm}
	+(-p_1\cdot\lambda)
	\left.
	\left(
	+\input{fig/st_s.tikz}
	+\input{fig/st_tp.tikz}
	\right)
	\right|_{\alpha_1\rightarrow\alpha_1+1} \notag\\
	&\hspace{60mm}
	+(-p_2'\cdot\lambda)
	\left.
	\left(
	+\input{fig/st_s.tikz}
	+\input{fig/st_tp.tikz}
	\right)
	\right|_{\alpha_2'\rightarrow\alpha_2'+1}
\end{align}
Using a formula for the hypergeometric function 
(see {\it e.g.} Eq.~15.8.4 in \cite{DLMF}),
\begin{align}
	&\pm\input{fig/st_s.tikz} +\input{fig/st_tp.tikz}  \notag\\
	&=
	\pm w^{1-\alpha_2-\alpha_1}
	\Gamma(-\alpha_2+1)\Gamma(-\alpha_1+1)\:
	\Freg{\alpha_2'}{-\alpha_2+1}{-\alpha_2-\alpha_1+2}{w} \notag\\
	&\hspace{30mm}
	+\Gamma(\alpha_2+\alpha_1+\alpha_2'-1)
	\Gamma(-\alpha_2+1)\:
	\Freg{\alpha_1}{\alpha_2+\alpha_1+\alpha_2'-1}{\alpha_1+\alpha_2'}{1-w} \notag\\
	&=
	\pm w^{1-\alpha_2-\alpha_1}
	\Gamma(-\alpha_2+1)\Gamma(-\alpha_1+1)\:
	\Freg{\alpha_2'}{-\alpha_2+1}{-\alpha_2-\alpha_1+2}{w} \notag\\
	&\hspace{5mm}
	+\Gamma(\alpha_2+\alpha_1+\alpha_2'-1) \Gamma(-\alpha_2+1)
	\frac{ \pi }{ \sin\pi(-\alpha_2-\alpha_1+1) } \notag\\
	&\hspace{5mm}\times\left[
	\frac{1}{ \Gamma(\alpha_2')\Gamma(-\alpha_2+1) }
	\Freg{\alpha_1}{\alpha_2+\alpha_1+\alpha_2'-1}{\alpha_2+\alpha_1}{w} \right.\notag\\
	&\hspace{60mm}\left.
	- \frac{ w^{-\alpha_2-\alpha_1+1} }{ \Gamma(\alpha_1)\Gamma(\alpha_2+\alpha_1+\alpha_2'-1) }
	\Freg{\alpha_2'}{-\alpha_2+1}{-\alpha_2-\alpha_1+2}{w}
	\right] \notag\\
	&\sim
	\pm w^{1-\alpha_2-\alpha_1}
	\Gamma(-\alpha_2+1)\Gamma(-\alpha_1+1)\:
	\Freg{\alpha_2'}{-\alpha_2+1}{-\alpha_2-\alpha_1+2}{w} \notag\\
	&\hspace{5mm}
	-\Gamma(\alpha_2+\alpha_1+\alpha_2'-1) \Gamma(-\alpha_2+1)
	\frac{ \pi }{ \sin\pi(-\alpha_2-\alpha_1+1) } \notag\\
	&\hspace{63mm}\times
	\frac{ w^{-\alpha_2-\alpha_1+1} }{ \Gamma(\alpha_1)\Gamma(\alpha_2+\alpha_1+\alpha_2'-1) }
	\Freg{\alpha_2'}{-\alpha_2+1}{-\alpha_2-\alpha_1+2}{w} \notag\\
	&=
	\left( \pm 1 -\frac{\sin\pi\alpha_1}{ \sin\pi(-\alpha_2-\alpha_1+1) } \right)
	\Gamma(-\alpha_2+1)\Gamma(-\alpha_1+1)\:
	w^{-\alpha_2-\alpha_1+1}
	\Freg{\alpha_2'}{-\alpha_2+1}{-\alpha_2-\alpha_1+2}{w} \notag\\
	&\sim
	\left( \pm 1 + (-1)^{-\alpha_2+1} \right)
	\input{fig/st_s.tikz}
\end{align}
When $N_a$ is even,
the integral $Z_a^{212'}(\alpha,p,\lambda;w)$ 
trivially vanishes.
Thus in the following computations,
we assume that $N_a$ is odd for any $a=1,\dots,J$.
Then
\begin{align}
	&Z_a^{212'}(\alpha,p,\lambda;w) \notag\\
	&\sim
	+(p_2\cdot\lambda)(+2) \left.
	\input{fig/st_s.tikz}
	\right|_{\alpha_2\rightarrow\alpha_2+1} \notag\\
	&\hspace{30mm}+
	(p_1\cdot\lambda)(-2) \left.
	\input{fig/st_s.tikz}
	\right|_{\alpha_1\rightarrow\alpha_1+1} \notag\\
	&\hspace{60mm}+
	(p_2'\cdot\lambda)(-2) \left.
	\input{fig/st_s.tikz}
	\right|_{\alpha_2'\rightarrow\alpha_2'+1}.
\end{align}

\subsection{Tachyon integral}

To proceed,
let us expand the hypergeometric function as
\begin{align}
	\input{fig/st_s.tikz}
	&= w^{-\alpha_2-\alpha_1+1}
	\Gamma(-\alpha_2+1)\Gamma(-\alpha_1+1)\:
	\Freg{\alpha_2'}{-\alpha_2+1}{-\alpha_2-\alpha_1+2}{w} \notag\\
	&= w^{-\alpha_2-\alpha_1+1}
	\sum_{k=0}^{\infty} c_k(\alpha_2,\alpha_1,\alpha_2')\: w^k.
\end{align}
Here we have defined
\begin{align}
	&c_k(\alpha_2,\alpha_1,\alpha_2') \notag\\
	&= \frac{ \Gamma(-\alpha_2+1)\Gamma(-\alpha_1+1) }{ \Gamma(-\alpha_2-\alpha_1+2) }
	\frac{ \Gamma(\alpha_2'+k) }{ \Gamma(\alpha_2') }
	\frac{ \Gamma(-\alpha_2+1+k) }{ \Gamma(-\alpha_2+1) }
	\frac{ \Gamma(-\alpha_2-\alpha_1+2) }{ \Gamma(-\alpha_2-\alpha_1+2+k) }
	\frac{1}{\Gamma(k+1)} \notag\\
	&\sim
	\frac{\pi}{\sin\pi\alpha_2}
	\frac{1}{\Gamma(k+1)\Gamma(\alpha_2-k)}
	\frac{\Gamma(\alpha_2-k+\alpha_1-1)}{\Gamma(\alpha_1)}
	\frac{\Gamma(\alpha_2'+k)}{\Gamma(\alpha_2')}.
\end{align}
Then
\begin{align}
	&Z_a^{212'}(\alpha,p,\lambda;w) \notag\\
	&\sim
	(p_2\cdot\lambda)(+2)\:
	w^{-\alpha_2-\alpha_1}
	c_k(\alpha_2+1,\alpha_1,\alpha_2')\: w^{k}
	+(p_1\cdot\lambda)(-2)\:
	w^{-\alpha_2-\alpha_1}
	c_k(\alpha_2,\alpha_1+1,\alpha_2')\: w^{k} \notag\\
	&\hspace{63mm}
	+(p_2'\cdot\lambda)(-2)\:
	w^{-\alpha_2-\alpha_1}
	\sum_{k=0}^{\infty}
	c_k(\alpha_2,\alpha_1,\alpha_2'+1)\:
	w^{k+1}.
	\label{eq:first_two_terms}
\end{align}
Now let us assume that
\begin{align}
	(p_1+p_2)\cdot\lambda = (p_2'+p_1')\cdot\lambda' = 0
\end{align}
for simplicity.
These conditions are satisfied, for example,
in the center-of-mass frame.
The first two terms in \eqref{eq:first_two_terms} are rearranged as
\begin{align}
	Z_a^{212'}(\alpha,p,\lambda;w)
	&\sim
	(p_2\cdot\lambda)(+2)\:
	w^{-\alpha_2-\alpha_1}
	\sum_{k=0}^{\alpha_2\sim N_a}
	c_k(\alpha_2+1,\alpha_1+1,\alpha_2')\:
	w^k \notag\\
	&\hspace{30mm}
	+(p_2'\cdot\lambda)(-2)\:
	w^{-\alpha_2-\alpha_1}
	\sum_{k=0}^{\alpha_2\sim N_a}
	c_{k-1}(\alpha_2,\alpha_1,\alpha_2'+1)\:
	w^{k}.
\end{align}
Similarly, by replacements
$2\leftrightarrow2',w\rightarrow1-w,\alpha\rightarrow\beta,\lambda\rightarrow\lambda'$,
we obtain
\begin{align}
	Z_b^{212'}(\beta,p,\lambda';w)
	&\sim
	(p_2'\cdot\lambda')(+2)\:
	(1-w)^{-\beta_2'-\beta_1}
	\sum_{l=0}^{\beta_2'\sim N_b'}
	c_l(\beta_2'+1,\beta_1+1,\beta_2)\:
	(1-w)^l \notag\\
	&\hspace{18mm}
	+(p_2\cdot\lambda')(-2)\:
	(1-w)^{-\beta_2'-\beta_1}
	\sum_{l=0}^{\beta_2'\sim N_b'}
	c_{l-1}(\beta_2',\beta_1,\beta_2+1)\:
	(1-w)^{l}.
\end{align}
Substituting these into \eqref{eq:hes_amp_st},
we can perform the integral over $w$.
Noting that
\begin{align}
	&p_1\cdot p_2 - \sum_{a=1}^{J}\alpha_2 - \sum_{a=1}^{J}\alpha_1
	= -2-s/2, \\
	&p_1\cdot p_2' - \sum_{b=1}^{J}\beta_2' - \sum_{b=1}^{J}\beta_1
	= -2-t/2,
\end{align}
the result is
\begin{align}
    \label{eq:amp_hes_tachyon_st-shifted}
	\mathcal{A}_{st}
	&\sim
	\sum_{ \{i_a=2,2'\} }
	\sum_{ \{j_b=2,2'\} }
	\sum_{ \{k_a=0\} }^{ \{N_a\} }
	\sum_{ \{l_b=0\} }^{ \{N_b'\} } \notag\\
	&\hspace{30mm}
	\left(
	\prod_{a=1}^{J}
	(p_{i_a}\cdot\lambda)\:
	c_{k_a}^{(i_a)}
	\right)
	\left(
	\prod_{b=1}^{J}
	(p_{j_b}\cdot\lambda')\:
	d_{l_b}^{(j_b)}
	\right)
	B(-\alpha(s)+k,-\alpha(t)+l)
\end{align}
where
\begin{align}
	&c_k^{(2)} = c_k(\alpha_2+1,\alpha_1+1,\alpha_2'), \\
	&c_k^{(2')} = -c_{k-1}(\alpha_2,\alpha_1,\alpha_2'+1), \\
	&d_l^{(2)} = -c_{l-1}(\beta_2',\beta_1,\beta_2+1), \\
	&d_l^{(2')} = c_l(\beta_2'+1,\beta_1+1,\beta_2),
\end{align}
and where
\begin{align}
	& B(a,b) = \frac{ \Gamma(a)\Gamma(b) }{ \Gamma(a+b) }, \quad
	\alpha(x) = 1+x/2, \\
	&k = \sum_{a=1}^{J} k_a, \quad
	l = \sum_{b=1}^{J} l_b.
\end{align}
Here ends the derivation of the HES-tachyon to HES-tachyon amplitude, given in \eqref{eq:amp_hes_tachyon_st-shifted}.

\subsection{Another expression for the amplitude}

In this subsection, for a consistency check of our amplitude formula
\eqref{eq:amp_hes_tachyon_st-shifted}, we study another expression 
of the amplitude.

Similar computations show that
\begin{align}
	&Z_b^{212'}(\beta,p,\lambda',w) \notag\\
	&\sim
	+(p_2\cdot\lambda')(+2)
	\left. \input{fig/st_b_sp.tikz} \right|_{\beta_2\rightarrow\beta_2+1}
	+(p_1\cdot\lambda')(+2)
	\left. \input{fig/st_b_sp.tikz} \right|_{\beta_1\rightarrow\beta_1+1} \notag\\
	&\hspace{80mm}
	+(p_2'\cdot\lambda')(+2)
	\left. \input{fig/st_b_sp.tikz} \right|_{\beta_2'\rightarrow\beta_2'+1} \notag\\
	&\sim
	+(p_2\cdot\lambda')(+2)
	\left(
	\left. \input{fig/st_b_sp.tikz} \right|_{\beta_2\rightarrow\beta_2+1}
	-\left. \input{fig/st_b_sp.tikz} \right|_{\beta_1\rightarrow\beta_1+1}
	\right) \notag\\
	&\hspace{80mm}
	+(p_2'\cdot\lambda')(+2)
	\left. \input{fig/st_b_sp.tikz} \right|_{\beta_2'\rightarrow\beta_2'+1}
\end{align}
Using a formula for the hypergeometric function
(see {\it e.g.} Eq.~15.8.1 in \cite{DLMF}),
\begin{align}
	&\left. \input{fig/st_b_sp.tikz} \right|_{\beta_2\rightarrow\beta_2+1}
	-\left. \input{fig/st_b_sp.tikz} \right|_{\beta_1\rightarrow\beta_1+1} \notag\\
	&=
	\Gamma(\beta_2'+\beta_1+\beta_2)\Gamma(-\beta_2'+1)
	\left[
	\Freg{\beta_1}{\beta_2'+\beta_1+\beta_2}{\beta_1+\beta_2+1}{w}
	-\Freg{\beta_1+1}{\beta_2'+\beta_1+\beta_2}{\beta_1+\beta_2+1}{w}
	\right] \notag\\
	&=
	\Gamma(\beta_2'+\beta_1+\beta_2+1)\Gamma(-\beta_2'+1)\:
	(-w)\:
	\Freg{\beta_1+1}{\beta_2'+\beta_1+\beta_2+1}{\beta_1+\beta_2+2}{w} \notag\\
	&=
	(1-w)^{-\beta_2'-\beta_1}\:
	\Gamma(\beta_2'+\beta_1+\beta_2+1)\Gamma(-\beta_2'+1)\:
	(-w)\:
	\Freg{\beta_2+1}{-\beta_2'+1}{\beta_1+\beta_2+2}{w} \notag\\
	&=
	(1-w)^{-\beta_2'-\beta_1}\:
	\Gamma(-\beta_1'+1)\Gamma(-\beta_2'+1)\:
	(-w)\:
	\Freg{\beta_2+1}{-\beta_2'+1}{-\beta_2'-\beta_1'+2}{w} \notag\\
	&=
	(1-w)^{-\beta_2'-\beta_1}\:
	(-w)\:
	\sum_{l=0}^{\infty}
	c_l(\beta_2',\beta_1',\beta_2+1)\:
	w^{l} \notag\\
	&=
	(1-w)^{-\beta_2'-\beta_1}\:
	\sum_{l=0}^{\beta_2'\sim N_b'}
	-c_{l-1}(\beta_2',\beta_1',\beta_2+1)\:
	w^{l}, \\
	&\left. \input{fig/st_b_sp.tikz} \right|_{\beta_2'\rightarrow\beta_2'+1} \notag\\
	&=
	\Gamma(\beta_2'+\beta_1+\beta_2)\Gamma(-\beta_2')\:
	\Freg{\beta_1}{\beta_2'+\beta_1+\beta_2}{\beta_1+\beta_2}{w} \notag\\
	&=
	(1-w)^{-\beta_2'-\beta_1}\:
	\Gamma(\beta_2'+\beta_1+\beta_2)\Gamma(-\beta_2')\:
	\Freg{\beta_2}{-\beta_2'}{\beta_1+\beta_2}{w} \notag\\
	&=
	(1-w)^{-\beta_2'-\beta_1}\:
	\Gamma(-\beta_1')\Gamma(-\beta_2')\:
	\Freg{\beta_2}{-\beta_2'}{-\beta_2'-\beta_1'}{w} \notag\\
	&=
	(1-w)^{-\beta_2'-\beta_1}\:
	\sum_{l=0}^{\beta_2'\sim N_b'}
	c_l(\beta_2'+1,\beta_1'+1,\beta_2)\:
	w^l.
\end{align}
Substituting these into \eqref{eq:hes_amp_st},
and noting that
\begin{align}
	&p_1\cdot p_2 - \sum_{a=1}^{J}\alpha_2 - \sum_{a=1}^{J}\alpha_1
	= -2-s/2, \\
	&p_1\cdot p_2' - \sum_{b=1}^{J}\beta_2' - \sum_{b=1}^{J}\beta_1
	= -2-t/2,
\end{align}
we can perform the integral over $w$.
The result is
\begin{align}
	\label{eq:amp_hes_tachyon_s-shifted}
	\mathcal{A}_{st}
	&\sim
	\sum_{ \{i_a=2,2'\} }
	\sum_{ \{j_b=2,2'\} }
	\sum_{ \{k_a=0\} }^{ \{N_a\} }
	\sum_{ \{l_b=0\} }^{ \{N_b'\} } \notag\\
	&\hspace{30mm}
	\left(
	\prod_{a=1}^{J}
	(p_{i_a}\cdot\lambda)\:
	c_{k_a}^{(i_a)}
	\right)
	\left(
	\prod_{b=1}^{J}
	(p_{j_b}\cdot\lambda')\:
	d_{l_b}^{(j_b)}
	\right)
	B(-\alpha(s)+k+l,-\alpha(t))
\end{align}
where
\begin{align}
	c_k^{(2)}  &= c_k(\alpha_2+1,\alpha_1+1,\alpha_2'), \\
	c_k^{(2')} &= -c_{k-1}(\alpha_2,\alpha_1,\alpha_2'+1), \\
	d_l^{(2)}   &= -c_{l-1}(\beta_2',\beta_1',\beta_2+1), \\
	d_l^{(2')}  &= c_l(\beta_2'+1,\beta_1'+1,\beta_2),
\end{align}
and where
\begin{align}
	&k = \sum_{a=1}^{J} k_a, \quad
	l = \sum_{b=1}^{J} l_b, \\
	& B(a,b) = \frac{ \Gamma(a)\Gamma(b) }{ \Gamma(a+b) }, \quad
	\alpha(x) = 1+x/2.
\end{align}
In spite of the superficially different appearance of the amplitude 
formula 
\eqref{eq:amp_hes_tachyon_st-shifted} and
the formula \eqref{eq:amp_hes_tachyon_s-shifted},
they are equivalent. 
The former depends on $\beta_1$
and both of the $s,t$-channels are shifted,
while the latter depends on $\beta_1'$
and only the $s$-channel is shifted.

\bibliographystyle{utphys}
\bibliography{ref}

\end{document}

%% file: fig/string_scattering.tikzstyles
\tikzstyle{disk}=[fill=white, draw=black, shape=circle, minimum size=10mm]
\tikzstyle{large disk}=[fill=white, draw=black, shape=circle, minimum size=20mm]

\tikzstyle{dotted}=[-, fill=none, densely dotted]

%% file: fig/2_tachyon_J_photon.tikz
\begin{tikzpicture}
	\begin{pgfonlayer}{nodelayer}
		\node [style=large disk] (0) at (0, 0) {};
		\node [style=none, label={below left:$p_1$}] (1) at (-4, -4) {};
		\node [style=disk] (2) at (-2.75, 2.75) {};
		\node [style=none, label={above left:$p_2$}] (3) at (-4, 4) {};
		\node [style=none, label={left:$-N_1q,\lambda$}] (4) at (-4.5, 3) {};
		\node [style=none, label={below left:$-N_Jq,\lambda$}] (5) at (-4, 1.5) {};
		\node [style=none, label={above right:$p_1'$}] (6) at (4, 4) {};
		\node [style=none, label={above right:$-N_J'q',\lambda'$}] (7) at (4, -1.5) {};
		\node [style=none, label={right:$-N_1'q',\lambda'$}] (8) at (4.5, -3) {};
		\node [style=none, label={below right:$p_2'$}] (9) at (4, -4) {};
		\node [style=disk] (10) at (2.75, -2.75) {};
		\node [style=none] (11) at (-6, 2.5) {};
		\node [style=none] (12) at (-5.75, 1.5) {};
		\node [style=none] (13) at (5.75, -1.5) {};
		\node [style=none] (14) at (6, -2.5) {};
		\node [style=none] (15) at (-1.25, 2.25) {$v$};
		\node [style=none] (17) at (2.25, -1.25) {$v'$};
	\end{pgfonlayer}
	\begin{pgfonlayer}{edgelayer}
		\draw (0) to (1.center);
		\draw (2) to (0);
		\draw (2) to (3.center);
		\draw [snake=snake, segment length=1.8mm] (2) to (4.center);
		\draw [snake=snake, segment length=1.8mm] (2) to (5.center);
		\draw (0) to (10);
		\draw [snake=snake, segment length=1.8mm] (10) to (7.center);
		\draw [snake=snake, segment length=1.8mm] (10) to (8.center);
		\draw (10) to (9.center);
		\draw (0) to (6.center);
		\draw [style=dotted, bend right=15, looseness=0.75] (11.center) to (12.center);
		\draw [style=dotted, bend right=15, looseness=0.75] (14.center) to (13.center);
	\end{pgfonlayer}
\end{tikzpicture}

%% file: fig/st_s.tikz
\begin{tikzpicture}
	\begin{pgfonlayer}{nodelayer}
		\node [style=none] (0) at (0, 0) {};
		\node [style=none, label={below right:$1$}] (1) at (1, -1) {};
		\node [style=none, label={below left:$2$}] (2) at (-1, -1) {};
		\node [style=none, label={above right:$2'$}] (3) at (1, 1) {};
		\node [style=none, label={above left:$1'$}] (4) at (-1, 1) {};
		\node [style=disk] (6) at (0, 0) {};
		\node [style=none, label={center:$\infty$}] (7) at (-0.5, 0.5) {};
		\node [style=none, label={center:$0$}] (8) at (-0.5, -0.5) {};
		\node [style=none, label={center:$w$}] (9) at (0.5, -0.5) {};
		\node [style=none, label={center:$1$}] (10) at (0.5, 0.5) {};
		\node [style=none, label={below:$a$}] (11) at (0, -1.5) {};
	\end{pgfonlayer}
	\begin{pgfonlayer}{edgelayer}
		\draw (0.center) to (1.center);
		\draw (0.center) to (2.center);
		\draw (0.center) to (3.center);
		\draw (0.center) to (4.center);
		\draw [snake=snake,segment length=1.3mm] (6) to (11.center);
	\end{pgfonlayer}
\end{tikzpicture}

%% file: fig/st_tp.tikz
\begin{tikzpicture}
	\begin{pgfonlayer}{nodelayer}
		\node [style=none] (0) at (0, 0) {};
		\node [style=none, label={below right:$1$}] (1) at (1, -1) {};
		\node [style=none, label={below left:$2$}] (2) at (-1, -1) {};
		\node [style=none, label={above right:$2'$}] (3) at (1, 1) {};
		\node [style=none, label={above left:$1'$}] (4) at (-1, 1) {};
		\node [style=disk] (6) at (0, 0) {};
		\node [style=none, label={center:$\infty$}] (7) at (-0.5, 0.5) {};
		\node [style=none, label={center:$0$}] (8) at (-0.5, -0.5) {};
		\node [style=none, label={center:$w$}] (9) at (0.5, -0.5) {};
		\node [style=none, label={center:$1$}] (10) at (0.5, 0.5) {};
		\node [style=none, label={left:$a$}] (11) at (-1.5, 0) {};
	\end{pgfonlayer}
	\begin{pgfonlayer}{edgelayer}
		\draw (0.center) to (1.center);
		\draw (0.center) to (2.center);
		\draw (0.center) to (3.center);
		\draw (0.center) to (4.center);
		\draw [snake=snake,segment length=1.3mm] (6) to (11.center);
	\end{pgfonlayer}
\end{tikzpicture}

%% file: fig/st_t.tikz
\begin{tikzpicture}
	\begin{pgfonlayer}{nodelayer}
		\node [style=none] (0) at (0, 0) {};
		\node [style=none, label={below right:$1$}] (1) at (1, -1) {};
		\node [style=none, label={below left:$2$}] (2) at (-1, -1) {};
		\node [style=none, label={above right:$2'$}] (3) at (1, 1) {};
		\node [style=none, label={above left:$1'$}] (4) at (-1, 1) {};
		\node [style=disk] (6) at (0, 0) {};
		\node [style=none, label={center:$\infty$}] (7) at (-0.5, 0.5) {};
		\node [style=none, label={center:$0$}] (8) at (-0.5, -0.5) {};
		\node [style=none, label={center:$w$}] (9) at (0.5, -0.5) {};
		\node [style=none, label={center:$1$}] (10) at (0.5, 0.5) {};
		\node [style=none, label={right:$a$}] (11) at (1.5, 0) {};
	\end{pgfonlayer}
	\begin{pgfonlayer}{edgelayer}
		\draw (0.center) to (1.center);
		\draw (0.center) to (2.center);
		\draw (0.center) to (3.center);
		\draw (0.center) to (4.center);
		\draw [snake=snake,segment length=1.3mm] (6) to (11.center);
	\end{pgfonlayer}
\end{tikzpicture}

%% file: fig/st_sp.tikz
\begin{tikzpicture}
	\begin{pgfonlayer}{nodelayer}
		\node [style=none] (0) at (0, 0) {};
		\node [style=none, label={below right:$1$}] (1) at (1, -1) {};
		\node [style=none, label={below left:$2$}] (2) at (-1, -1) {};
		\node [style=none, label={above right:$2'$}] (3) at (1, 1) {};
		\node [style=none, label={above left:$1'$}] (4) at (-1, 1) {};
		\node [style=disk] (6) at (0, 0) {};
		\node [style=none, label={center:$\infty$}] (7) at (-0.5, 0.5) {};
		\node [style=none, label={center:$0$}] (8) at (-0.5, -0.5) {};
		\node [style=none, label={center:$w$}] (9) at (0.5, -0.5) {};
		\node [style=none, label={center:$1$}] (10) at (0.5, 0.5) {};
		\node [style=none, label={above:$a$}] (11) at (0, 1.5) {};
	\end{pgfonlayer}
	\begin{pgfonlayer}{edgelayer}
		\draw (0.center) to (1.center);
		\draw (0.center) to (2.center);
		\draw (0.center) to (3.center);
		\draw (0.center) to (4.center);
		\draw [snake=snake,segment length=1.3mm] (6) to (11.center);
	\end{pgfonlayer}
\end{tikzpicture}

%% file: fig/st_s-s.tikz
\begin{tikzpicture}
	\begin{pgfonlayer}{nodelayer}
		\node [style=none] (0) at (0, 0) {};
		\node [style=none, label={below right:$1$}] (1) at (1, -1) {};
		\node [style=none, label={below left:$2$}] (2) at (-1, -1) {};
		\node [style=none, label={above right:$2'$}] (3) at (1, 1) {};
		\node [style=none, label={above left:$1'$}] (4) at (-1, 1) {};
		\node [style=disk] (6) at (0, 0) {};
		\node [style=none, label={center:$\infty$}] (7) at (-0.5, 0.5) {};
		\node [style=none, label={center:$0$}] (8) at (-0.5, -0.5) {};
		\node [style=none, label={center:$w$}] (9) at (0.5, -0.5) {};
		\node [style=none, label={center:$1$}] (10) at (0.5, 0.5) {};
		\node [style=none, label={below:$a$}] (11) at (0, -1.5) {};
		\node [style=none] (12) at (-1.5, -2) {};
		\node [style=none] (13) at (-1.5, -2.5) {};
		\node [style=none] (14) at (0, -2.5) {};
		\node [style=none] (15) at (0, -2.25) {};
	\end{pgfonlayer}
	\begin{pgfonlayer}{edgelayer}
		\draw (0.center) to (1.center);
		\draw (0.center) to (2.center);
		\draw (0.center) to (3.center);
		\draw (0.center) to (4.center);
		\draw [snake=snake, segment length=1.3mm] (6) to (11.center);
		\draw [->] (13.center) to (12.center);
		\draw (13.center) to (14.center);
		\draw (14.center) to (15.center);
	\end{pgfonlayer}
\end{tikzpicture}

%% file: fig/st_s-n.tikz
\begin{tikzpicture}
	\begin{pgfonlayer}{nodelayer}
		\node [style=none] (0) at (0, 0) {};
		\node [style=none, label={below right:$1$}] (1) at (1, -1) {};
		\node [style=none, label={below left:$2$}] (2) at (-1, -1) {};
		\node [style=none, label={above right:$2'$}] (3) at (1, 1) {};
		\node [style=none, label={above left:$1'$}] (4) at (-1, 1) {};
		\node [style=disk] (6) at (0, 0) {};
		\node [style=none, label={center:$\infty$}] (7) at (-0.5, 0.5) {};
		\node [style=none, label={center:$0$}] (8) at (-0.5, -0.5) {};
		\node [style=none, label={center:$w$}] (9) at (0.5, -0.5) {};
		\node [style=none, label={center:$1$}] (10) at (0.5, 0.5) {};
		\node [style=none, label={below:$a$}] (11) at (0, -1.5) {};
		\node [style=none] (12) at (1.5, -2) {};
		\node [style=none] (13) at (1.5, -2.5) {};
		\node [style=none] (14) at (0, -2.5) {};
		\node [style=none] (15) at (0, -2.25) {};
	\end{pgfonlayer}
	\begin{pgfonlayer}{edgelayer}
		\draw (0.center) to (1.center);
		\draw (0.center) to (2.center);
		\draw (0.center) to (3.center);
		\draw (0.center) to (4.center);
		\draw [snake=snake, segment length=1.3mm] (6) to (11.center);
		\draw [->] (13.center) to (12.center);
		\draw (13.center) to (14.center);
		\draw (14.center) to (15.center);
	\end{pgfonlayer}
\end{tikzpicture}

%% file: fig/st_s-t.tikz
\begin{tikzpicture}
	\begin{pgfonlayer}{nodelayer}
		\node [style=none] (0) at (0, 0) {};
		\node [style=none, label={below right:$1$}] (1) at (1, -1) {};
		\node [style=none, label={below left:$2$}] (2) at (-1, -1) {};
		\node [style=none, label={above right:$2'$}] (3) at (1, 1) {};
		\node [style=none, label={above left:$1'$}] (4) at (-1, 1) {};
		\node [style=disk] (6) at (0, 0) {};
		\node [style=none, label={center:$\infty$}] (7) at (-0.5, 0.5) {};
		\node [style=none, label={center:$0$}] (8) at (-0.5, -0.5) {};
		\node [style=none, label={center:$w$}] (9) at (0.5, -0.5) {};
		\node [style=none, label={center:$1$}] (10) at (0.5, 0.5) {};
		\node [style=none, label={below:$a$}] (11) at (0, -1.5) {};
		\node [style=none] (12) at (2, 1.5) {};
		\node [style=none] (15) at (0, -2.25) {};
		\node [style=none] (16) at (2.5, 1.5) {};
		\node [style=none] (17) at (0, -2.5) {};
		\node [style=none] (18) at (2.5, -2.5) {};
	\end{pgfonlayer}
	\begin{pgfonlayer}{edgelayer}
		\draw (0.center) to (1.center);
		\draw (0.center) to (2.center);
		\draw (0.center) to (3.center);
		\draw (0.center) to (4.center);
		\draw [snake=snake, segment length=1.3mm] (6) to (11.center);
		\draw (15.center) to (17.center);
		\draw (17.center) to (18.center);
		\draw (18.center) to (16.center);
		\draw [->] (16.center) to (12.center);
	\end{pgfonlayer}
\end{tikzpicture}

%% file: fig/st_tp-s.tikz
\begin{tikzpicture}
	\begin{pgfonlayer}{nodelayer}
		\node [style=none] (0) at (0, 0) {};
		\node [style=none, label={below right:$1$}] (1) at (1, -1) {};
		\node [style=none, label={below left:$2$}] (2) at (-1, -1) {};
		\node [style=none, label={above right:$2'$}] (3) at (1, 1) {};
		\node [style=none, label={above left:$1'$}] (4) at (-1, 1) {};
		\node [style=disk] (6) at (0, 0) {};
		\node [style=none, label={center:$\infty$}] (7) at (-0.5, 0.5) {};
		\node [style=none, label={center:$0$}] (8) at (-0.5, -0.5) {};
		\node [style=none, label={center:$w$}] (9) at (0.5, -0.5) {};
		\node [style=none, label={center:$1$}] (10) at (0.5, 0.5) {};
		\node [style=none, label={left:$a$}] (11) at (-1.5, 0) {};
		\node [style=none] (12) at (-2, -1.5) {};
		\node [style=none] (13) at (-2.5, -1.5) {};
		\node [style=none] (14) at (-2.5, 0) {};
		\node [style=none] (15) at (-2.25, 0) {};
	\end{pgfonlayer}
	\begin{pgfonlayer}{edgelayer}
		\draw (0.center) to (1.center);
		\draw (0.center) to (2.center);
		\draw (0.center) to (3.center);
		\draw (0.center) to (4.center);
		\draw [snake=snake, segment length=1.3mm] (6) to (11.center);
		\draw [->] (13.center) to (12.center);
		\draw (13.center) to (14.center);
		\draw (14.center) to (15.center);
	\end{pgfonlayer}
\end{tikzpicture}

%% file: fig/st_tp-n.tikz
\begin{tikzpicture}
	\begin{pgfonlayer}{nodelayer}
		\node [style=none] (0) at (0, 0) {};
		\node [style=none, label={below right:$1$}] (1) at (1, -1) {};
		\node [style=none, label={below left:$2$}] (2) at (-1, -1) {};
		\node [style=none, label={above right:$2'$}] (3) at (1, 1) {};
		\node [style=none, label={above left:$1'$}] (4) at (-1, 1) {};
		\node [style=disk] (6) at (0, 0) {};
		\node [style=none, label={center:$\infty$}] (7) at (-0.5, 0.5) {};
		\node [style=none, label={center:$0$}] (8) at (-0.5, -0.5) {};
		\node [style=none, label={center:$w$}] (9) at (0.5, -0.5) {};
		\node [style=none, label={center:$1$}] (10) at (0.5, 0.5) {};
		\node [style=none, label={left:$a$}] (11) at (-1.5, 0) {};
		\node [style=none] (12) at (1.5, -2) {};
		\node [style=none] (13) at (1.5, -2.5) {};
		\node [style=none] (15) at (-2.25, 0) {};
		\node [style=none] (16) at (-2.5, 0) {};
		\node [style=none] (17) at (-2.5, -2.5) {};
	\end{pgfonlayer}
	\begin{pgfonlayer}{edgelayer}
		\draw (0.center) to (1.center);
		\draw (0.center) to (2.center);
		\draw (0.center) to (3.center);
		\draw (0.center) to (4.center);
		\draw [snake=snake, segment length=1.3mm] (6) to (11.center);
		\draw [->] (13.center) to (12.center);
		\draw (15.center) to (16.center);
		\draw (16.center) to (17.center);
		\draw (17.center) to (13.center);
	\end{pgfonlayer}
\end{tikzpicture}

%% file: fig/st_tp-t.tikz
\begin{tikzpicture}
	\begin{pgfonlayer}{nodelayer}
		\node [style=none] (0) at (0, 0) {};
		\node [style=none, label={below right:$1$}] (1) at (1, -1) {};
		\node [style=none, label={below left:$2$}] (2) at (-1, -1) {};
		\node [style=none, label={above right:$2'$}] (3) at (1, 1) {};
		\node [style=none, label={above left:$1'$}] (4) at (-1, 1) {};
		\node [style=disk] (6) at (0, 0) {};
		\node [style=none, label={center:$\infty$}] (7) at (-0.5, 0.5) {};
		\node [style=none, label={center:$0$}] (8) at (-0.5, -0.5) {};
		\node [style=none, label={center:$w$}] (9) at (0.5, -0.5) {};
		\node [style=none, label={center:$1$}] (10) at (0.5, 0.5) {};
		\node [style=none, label={left:$a$}] (11) at (-1.5, 0) {};
		\node [style=none] (12) at (2, 1.5) {};
		\node [style=none] (15) at (-2.25, 0) {};
		\node [style=none] (16) at (2.5, 1.5) {};
		\node [style=none] (18) at (2.5, -2.5) {};
		\node [style=none] (19) at (-2.5, 0) {};
		\node [style=none] (20) at (-2.5, -2.5) {};
	\end{pgfonlayer}
	\begin{pgfonlayer}{edgelayer}
		\draw (0.center) to (1.center);
		\draw (0.center) to (2.center);
		\draw (0.center) to (3.center);
		\draw (0.center) to (4.center);
		\draw [snake=snake, segment length=1.3mm] (6) to (11.center);
		\draw (18.center) to (16.center);
		\draw [->] (16.center) to (12.center);
		\draw (15.center) to (19.center);
		\draw (19.center) to (20.center);
		\draw (20.center) to (18.center);
	\end{pgfonlayer}
\end{tikzpicture}

%% file: fig/st_b_sp.tikz
\begin{tikzpicture}
	\begin{pgfonlayer}{nodelayer}
		\node [style=none] (0) at (0, 0) {};
		\node [style=none, label={below right:$1$}] (1) at (1, -1) {};
		\node [style=none, label={below left:$2$}] (2) at (-1, -1) {};
		\node [style=none, label={above right:$2'$}] (3) at (1, 1) {};
		\node [style=none, label={above left:$1'$}] (4) at (-1, 1) {};
		\node [style=disk] (6) at (0, 0) {};
		\node [style=none, label={center:$\infty$}] (7) at (-0.5, 0.5) {};
		\node [style=none, label={center:$0$}] (8) at (-0.5, -0.5) {};
		\node [style=none, label={center:$w$}] (9) at (0.5, -0.5) {};
		\node [style=none, label={center:$1$}] (10) at (0.5, 0.5) {};
		\node [style=none, label={above:$b$}] (11) at (0, 1.5) {};
	\end{pgfonlayer}
	\begin{pgfonlayer}{edgelayer}
		\draw (0.center) to (1.center);
		\draw (0.center) to (2.center);
		\draw (0.center) to (3.center);
		\draw (0.center) to (4.center);
		\draw [snake=snake,segment length=1.3mm] (6) to (11.center);
	\end{pgfonlayer}
\end{tikzpicture}

%% file: main.bbl
\providecommand{\href}[2]{#2}\begingroup\raggedright\begin{thebibliography}{10}

\bibitem{Horowitz:1996nw}
G.~T. Horowitz and J.~Polchinski, ``{A Correspondence principle for black holes
  and strings},'' \href{http://dx.doi.org/10.1103/PhysRevD.55.6189}{{\em Phys.
  Rev. D} {\bfseries 55} (1997) 6189--6197},
  \href{http://arxiv.org/abs/hep-th/9612146}{{\ttfamily arXiv:hep-th/9612146}}.

\bibitem{Horowitz:1997jc}
G.~T. Horowitz and J.~Polchinski, ``{Selfgravitating fundamental strings},''
  \href{http://dx.doi.org/10.1103/PhysRevD.57.2557}{{\em Phys. Rev. D}
  {\bfseries 57} (1998) 2557--2563},
  \href{http://arxiv.org/abs/hep-th/9707170}{{\ttfamily arXiv:hep-th/9707170}}.

\bibitem{Amati:1999fv}
D.~Amati and J.~G. Russo, ``{Fundamental strings as black bodies},''
  \href{http://dx.doi.org/10.1016/S0370-2693(99)00375-5}{{\em Phys. Lett. B}
  {\bfseries 454} (1999) 207--212},
  \href{http://arxiv.org/abs/hep-th/9901092}{{\ttfamily arXiv:hep-th/9901092}}.

\bibitem{Chen:2021dsw}
Y.~Chen, J.~Maldacena, and E.~Witten, ``{On the black hole/string
  transition},'' \href{http://arxiv.org/abs/2109.08563}{{\ttfamily
  arXiv:2109.08563 [hep-th]}}.

\bibitem{Maldacena:1997re}
J.~M. Maldacena, ``{The Large N limit of superconformal field theories and
  supergravity},'' \href{http://dx.doi.org/10.1023/A:1026654312961}{{\em Adv.
  Theor. Math. Phys.} {\bfseries 2} (1998) 231--252},
  \href{http://arxiv.org/abs/hep-th/9711200}{{\ttfamily arXiv:hep-th/9711200}}.

\bibitem{Shenker:2013pqa}
S.~H. Shenker and D.~Stanford, ``{Black holes and the butterfly effect},''
  \href{http://dx.doi.org/10.1007/JHEP03(2014)067}{{\em JHEP} {\bfseries 03}
  (2014) 067}, \href{http://arxiv.org/abs/1306.0622}{{\ttfamily arXiv:1306.0622
  [hep-th]}}.

\bibitem{Shenker:2013yza}
S.~H. Shenker and D.~Stanford, ``{Multiple Shocks},''
  \href{http://dx.doi.org/10.1007/JHEP12(2014)046}{{\em JHEP} {\bfseries 12}
  (2014) 046}, \href{http://arxiv.org/abs/1312.3296}{{\ttfamily arXiv:1312.3296
  [hep-th]}}.

\bibitem{Maldacena:2015waa}
J.~Maldacena, S.~H. Shenker, and D.~Stanford, ``{A bound on chaos},''
  \href{http://dx.doi.org/10.1007/JHEP08(2016)106}{{\em JHEP} {\bfseries 08}
  (2016) 106}, \href{http://arxiv.org/abs/1503.01409}{{\ttfamily
  arXiv:1503.01409 [hep-th]}}.

\bibitem{Gross:2021gsj}
D.~J. Gross and V.~Rosenhaus, ``{Chaotic scattering of highly excited
  strings},'' \href{http://dx.doi.org/10.1007/JHEP05(2021)048}{{\em JHEP}
  {\bfseries 05} (2021) 048}, \href{http://arxiv.org/abs/2103.15301}{{\ttfamily
  arXiv:2103.15301 [hep-th]}}.

\bibitem{Rosenhaus:2020tmv}
V.~Rosenhaus, ``{Chaos in the Quantum Field Theory S-Matrix},''
  \href{http://dx.doi.org/10.1103/PhysRevLett.127.021601}{{\em Phys. Rev.
  Lett.} {\bfseries 127} no.~2, (2021) 021601},
  \href{http://arxiv.org/abs/2003.07381}{{\ttfamily arXiv:2003.07381
  [hep-th]}}.

\bibitem{Rosenhaus:2021xhm}
V.~Rosenhaus, ``{Chaos in a Many-String Scattering Amplitude},''
  \href{http://dx.doi.org/10.1103/PhysRevLett.129.031601}{{\em Phys. Rev.
  Lett.} {\bfseries 129} no.~3, (2022) 031601},
  \href{http://arxiv.org/abs/2112.10269}{{\ttfamily arXiv:2112.10269
  [hep-th]}}.

\bibitem{Fukushima:2022lsd}
O.~Fukushima and K.~Yoshida, ``{Chaotic instability in the BFSS matrix
  model},'' \href{http://arxiv.org/abs/2204.06391}{{\ttfamily arXiv:2204.06391
  [hep-th]}}.

\bibitem{Firrotta:2022cku}
M.~Firrotta and V.~Rosenhaus, ``{Photon emission from an excited string},''
  \href{http://arxiv.org/abs/2207.01641}{{\ttfamily arXiv:2207.01641
  [hep-th]}}.

\bibitem{Bianchi:2022mhs}
M.~Bianchi, M.~Firrotta, J.~Sonnenschein, and D.~Weissman, ``{A measure for
  chaotic scattering amplitudes},''
  \href{http://arxiv.org/abs/2207.13112}{{\ttfamily arXiv:2207.13112
  [hep-th]}}.

\bibitem{Hashimoto:2022ugt}
K.~Hashimoto, Y.~Matsuo, and T.~Yoda, ``{String is a double slit},''
  \href{http://arxiv.org/abs/2206.10951}{{\ttfamily arXiv:2206.10951
  [hep-th]}}.

\bibitem{seoane2012new}
J.~M. Seoane and M.~A. Sanju{\'a}n, ``New developments in classical chaotic
  scattering,'' {\em Reports on Progress in Physics} {\bfseries 76} no.~1,
  (2012) 016001.

\bibitem{DelGiudice:1971yjh}
E.~Del~Giudice, P.~Di~Vecchia, and S.~Fubini, ``{General properties of the dual
  resonance model},''
  \href{http://dx.doi.org/10.1016/0003-4916(72)90272-2}{{\em Annals Phys.}
  {\bfseries 70} (1972) 378--398}.

\bibitem{bleher1990bifurcation}
S.~Bleher, C.~Grebogi, and E.~Ott, ``Bifurcation to chaotic scattering,'' {\em
  Physica D: Nonlinear Phenomena} {\bfseries 46} no.~1, (1990) 87--121.

\bibitem{Bianchi:2019ywd}
M.~Bianchi and M.~Firrotta, ``{DDF operators, open string coherent states and
  their scattering amplitudes},''
  \href{http://dx.doi.org/10.1016/j.nuclphysb.2020.114943}{{\em Nucl. Phys. B}
  {\bfseries 952} (2020) 114943},
  \href{http://arxiv.org/abs/1902.07016}{{\ttfamily arXiv:1902.07016
  [hep-th]}}.

\bibitem{DLMF}
``Degital library of mathematical functions, chapter 15 hypergeometric
  function.'' \url{https://dlmf.nist.gov/15.8}.
\newblock Accessed: 2022-08-10.

\end{thebibliography}\endgroup
